\documentclass[aip,pof,preprint]{revtex4-1}

\usepackage{upgreek}
\usepackage{graphicx}
\usepackage{subfigure}
\usepackage{amsfonts,amssymb,amsbsy,amsmath,amsthm}
\usepackage{tikz}
\usepackage{bm}
\usepackage{epstopdf}
\usepackage{dcolumn}
\usepackage[utf8]{inputenc}
\usepackage[T1]{fontenc}

\begin{document}

\title{Eulerian formulation for the triboelectric charging of polydisperse powder flows}

\author{Metin Zeybek}
\affiliation{Physikalisch-Technische Bundesanstalt~(PTB), Bundesallee 100, 38116 Braunschweig, Germany}
\author{Holger Grosshans}
\email[]{holger.grosshans@ptb.de}
\affiliation{Physikalisch-Technische Bundesanstalt~(PTB), Bundesallee 100, 38116 Braunschweig, Germany}
\affiliation{Institute of Apparatus- and Environmental Technology, Otto von Guericke University of Magdeburg, Universitätsplatz 2, 39106 Magdeburg, Germany}

\date{\today}

\begin{abstract}
We propose an Eulerian formulation for the triboelectric charging of powder flows.
Recently, continuum descriptions of the charging of monodisperse and bidisperse powder appeared.
Now, we propose the first Eulerian formulation for this type of flows that can fully account for polydisperse particle size distributions.
To this end, the joint particle size, velocity, and charge distribution function is solved through the direct quadrature method of moments.
The newly established approach includes the uptake of electrostatic charge by the particles when contacting a solid wall and the exchange of charge in-between particles during collisions.
\textcolor{black}{The electrostatic field induced by the charge of the particles and the drag forces of the surrounding gas affect the particle dynamics.}
Test cases of two-dimensional steady channel flows demonstrate the charging of the flow from the walls towards the center and in downstream direction.
Further, the method predicts the effect of variations of the particle size and initial velocity distribution and the charge diffusion coefficient.
The ability to handle polydispersity is a step towards the simulation of the electric charge build-up of real powder flows in full-scale technical applications.
\end{abstract}

\maketitle

\section{\label{Introduction}Introduction}

"Triboelectric charging" or simply "tribo-charging" is the phenomenon of charge exchange between solid surfaces, such as particles, when brought into contact.
Over the years, triboelectric charging found its way to several technical applications, such as electrostatic separation of insulators\cite{4305317, 1518821} or electrophotography\cite{schein1998recent}.
Aside from its industrial significance, examples of this phenomenon can be encountered in nature.
For example, strong electrostatic interactions during volcanic eruptions,\cite{miura2002measurements} aeolian transport of grains,\cite{kamra1972physical} and the formation of sharp-edged "razorbacks" observed on Mars\cite{shinbrot2006triboelectrification} can be traced back to triboelectric charging.

Also, electrostatic charge build-up is associated with numerous industrial accidents.
During the pneumatic transportation process, particles experience many collisions with each other and pipe walls.
These collisions, which go along with charge transfer, lead to an electrically charged particle-laden flow.
This might, depending on the specific transportation conditions, cause several problems.
Among these, electrostatic adhesion forces between the charged powder and pipe walls have been an important concern for retaining a steady flow.\cite{joseph1983vertical, nifuku1989static, adhiwidjaja2000simultaneous, yao2004electrostatics,Gro20c}
However, the consequences may be far more severe when an excessive charge accumulation brings about a sudden discharge.
This sudden release can lead to disastrous outcomes such as fires or dust explosions.\cite{nifuku2003study, Glor03,Glor09}
 
Its complex nature and far-fetching impact brought triboelectric charging into the limelight of several research projects.
Throughout the years, many experimental studies have been carried out to highlight the effects of various parameters on the charging of the powders.\cite{masuda1976static,artana1997contribution, tanoue1999effect, nifuku2003study, nomura2003environment, watano2003numerical, watano2006mechanism, fath2013electrostatic, Gro17b, Gro20c,TAGH}
These studies have shown to be, although highly contributive, not conclusive due to uncertainties involved in an experimental environment.
The most notable uncertainty might be related to the initial and boundary conditions of the particulate phase.

Numerical studies, on the other hand, offer great convenience to overcome the uncertainties involved in the prescription of initial and boundary conditions.
To this end, Eulerian-Lagrangian approaches are most popular in recent simulations.\citep{Gro20b,ZHOU2021,LIU2020}
Therein, each particle is tracked individually which makes the prescription of properties such as the particle size distribution straightforward.
This motivated the model developed by \citet{hassani2013numerical} to investigate the effect of electrostatic forces on the hydrodynamics of fluidized beds.
Their results indicated a decrease in the bubble sizes for the charged particles and an increase in the bed voidage, which was compared to the experimental works of \citet{bokkers2004mixing}.
\citet{laurentie2013discrete} also used an Eulerian-Lagrangian model to examine the triboelectric charging taking place in vertically-vibrated beds of granular plastics.
To study the electrostatic charge build-up during pneumatic conveying, \citet{Gro16a} developed a four-way coupled large-eddy simulation approach.
Finally, \citet{grosshans2017direct} implemented a direct numerical simulation approach to explore the interplay between the different physical mechanisms underlying particle electrification in a turbulent channel flow.
The study also underlined the significance of the particle Stokes number on charge build-up.

While the Eulerian-Lagrangian approach made its way in numerical studies of laboratory-scaled systems, the heavy burden of individual particle tracking limits the number of particles that can be computed simultaneously.
Contrary, the description of the particulate phase in an Eulerian framework opens the possibility to handle complete technical flows consisting of a vast amount of particles.
In the Eulerian description, the particulate phase is treated as a continuum whose properties are spatially and temporally averaged.

While the Eulerian-Eulerian approach is popular for general powder flow simulations, only recently a few studies appeared where it was employed to the charge generation of particle-laden flows.
In their work, \citet{kolehmainen2018eulerian} developed a two-fluid model derived from the Boltzmann equation for the number density function of particles.
Further, the Kinetic Theory of Granular Flow (KTGF)\cite{jenkins1983theory} was applied for the closure of the collision operator accounting for the changes in the number density function due to particle-particle collisions.
The study also proposes a set of boundary conditions for the particle charge at conducting walls.
To account for the charge exchange, the electrical model of \citet{laurentie2013discrete} is adopted.
The effect of electrostatic forces on the particles, which emerge when charge builds up, is included in their mathematical model.
The results highlight the effects of triboelectric charging and particle motion on charge transport, in which charge diffusion through the random motion of particles was found to be more significant than triboelectric diffusion.
Compared to the discrete element model, the proposed model performed fairly well for high solid phase volume fractions.
Deviations found on the lower volume fraction regime were attributed to the Maxwellian assumptions of particle charge and velocity.
\citet{ray2019euler} developed another two-fluid model and validated it against a lab-scale experiment of a fluidized bed of polyethylene particles.
In the proposed approach, the charge separation mechanism follows the procedures of \citet{matsusaka2000electrification} for the charge transfer at the conducting walls and of \citet{johnson1987frictional} for the boundary condition.
Similar to \citet{kolehmainen2018eulerian}, the model favors the KTGF \cite{jenkins1983theory} for the particle-particle encounters and adopts the electrical model proposed by \citet{laurentie2013discrete}.
Also, \citet{mont20} recently presented a new formulation to compute electrostatic charging of mono-disperse particles in Eulerian framework.

A common challenge in Eulerian modeling is the difficulty to account for the size distribution of the dispersed phase.
As elaborated in the recent review of \citet{Chow21}, this issue is probably the most important outstanding problem in the Eulerian modeling of charged particulates.
Thus, both aforementioned works are limited to monodisperse powder flows.
In this light the recent numerical development by \citet{ray2020eulerian} represents an important step forward:
they revised earlier two-fluid models by adopting the kinetic theory for bi-disperse granular flows based on the works of \citet{jenkins1987balance} for the closure of the collision operator.
Using the new model, they simulated steady-state flows of bi-disperse polyethylene particles in a one-dimensional domain.
They did not consider advection or moment transport, but electrostatic forces.
The results produced a bi-polar charge distribution for the particles of different sizes.
The study also highlights the necessity of a poly-disperse model, where broad size distributions are expected.

An Eulerian model for the triboelectric charging of polydisperse powder is not existing yet.
However, in another field, namely the simulation of sprays, the issue of the  poly-dispersity of the droplets was addressed by the development of moment-based methods.\citep{Mar03,Mcg97}
One variant of moment-based methods is the direct quadrature method of moments~(DQMOM)~\citep{Mar05} which turned out to be an accurate and efficient approach to compute sprays\citep{Fox08} and spray drying processes~\citep{Gro14f,Gro15b}.
The feature that distinguishes DQMOM from other moment-based methods is the direct transportation of the weights and abscissas of the distribution functions in terms of a discrete number of physical quantities.
So far, the electrical charging of particles was not integrated in DQMOM.

In this work, we propose an Eulerian approach adopting a DQMOM method with quadrature-based approximation for the closure of the moments.
This approach, aside from introducing the polydispersity to the numerical studies of triboelectric charging of powders, also considers the momentum and charge transfer between the particles and other surfaces.
The mathematical model is outlined in Section~II and its derivation is detailed in the Appendix.
In Section~III the conditions of the cases for which we tested our new model are given.
The results are presented and discussed in Section~IV followed by the concluding remarks in the last section.

\section{DQMOM formulation for electrically charging powder flow}
\label{DQMOM}

The particulate phase is considered to behave as a continuum characterized by the particle number density function,
\begin{equation}
\label{eq:particlePDF}
f=f(\bm{u},r,q;\bm{x},t) \, .
\end{equation}
This expression gives the probable number density of particles at time instance $t$ located at $\bm x$ having a velocity of $\bm{u}$, a radius of $r$, and carrying an electrical charge of $q$. 
We formulate the evolution of $f$ in terms of a population balance equation, analogous to Williams' spray equation\citep{Wil58} for droplets and to Boltzmann's equation for molecules, 
\begin{equation}
\label{eq:williams}
\dfrac{\partial f}{\partial t} + \dfrac{\partial (\bm{u}f)}{\partial \bm{x}} = - \bm{F}_f - q_f \, .
\end{equation}
Herein, the term $\bm{F}_f = \partial (\sum \dot{\bm{u}} \, f) /\partial \bm{u}$ represents the sum of the forces acting on the particles that affect $f$.
The novelty of our formulation is the term
\begin{equation}
\label{eq:qf}
q_f = \partial (\dot{q} \, f) / \partial q + \nu_q \dfrac{\partial \left(f \partial^2 q / \partial {\bm x}^2 \right)}{\partial q}
\end{equation}
which represents the change of the particle number density function, $f$, due to the change of the particle charge. 
The first term on the right and side of the above equation represents the effect of the temporal change of charge due to contact with a wall or other particles.
The second term relates to the diffusion of charge in space, either via the motion of the particles that carry charge or spatial charge transport through collisions with other particles.\cite{Gro18c}
The charge diffusion coefficient, $\nu_q$, implicitely includes both these effects.

The DQMOM formulation of the left hand side of equation~(\ref{eq:williams}) is given in the section below, followed by the formulation of the right hand side in Section~\ref{sec:rhs}.
Afterward, $q_f$ and the other source terms are detailed in Section~\ref{sec:source}.

\subsection{Spatial and temporal evolution}



According to the methodology of DQMOM, $f$ is quadrature-based approximated as the sum over $N$ nodes of the product of weighted Dirac-delta distribution ($\delta$) of the particles' velocities and charges, i.e.,
\begin{equation}
\label{qadratureapprox}
f = \sum^N_{n=1} w_n \delta(r-r_n) \delta(\bm{u}-\bm{u}_n) \delta(q-q_n) \, .
\end{equation}
In this equation, $w$ represents the particle weights or number densities.
For conciseness we use in the above and the following equations the short notation
\begin{equation}
\delta(\bm{u}-\bm{u}_n) = \delta(u_1-u_{\rm 1,n}) \delta(u_2-u_{\rm 2,n}) \delta(u_3-u_{\rm 3,n}) \, .
\end{equation}
Further, it is noted that we develop the new method for an arbitrary number of nodes $N$, instationary flows, and for three-dimensional domains.
However, the simulations which we present in this paper regard the case of $N=3$ and steady, two-dimensional flows.

Introducing the quadrature approximation into equation~(\ref{eq:williams}) leads to set of transport equations for the particle weights, and the abscissa of the radius, velocity,\citep{Fox08,gopireddy2014modeling} and charge distribution functions which read
\begin{align}
\label{eq:lhs41}
\dfrac{\partial (w_n)}{\partial t} + \dfrac{\partial (w_n \bm{u}_n)}{\partial \bm{x}} &= a_n
\\
\label{eq:lhs42}
\dfrac{\partial (w_n r_n)}{\partial t} + \dfrac{\partial (w_n \bm{u}_n r_n)}{\partial \bm{x}} &= b_n
\\
\label{eq:lhs43}
\dfrac{\partial (w_n r_n \bm{u}_n)}{\partial t} + \dfrac{\partial (w_n \bm{u}_n r_n \bm{u}_n)}{\partial \bm{x}} &= \bm{c}_n
\\
\label{eq:lhs44}
\dfrac{\partial (w_n r_n q_n)}{\partial t} + \dfrac{\partial (w_n \bm{u}_n r_n q_n)}{\partial \bm{x}} &= d_n
\end{align}
Thus, the fields for $w_n$, $r_n$, $\bm{u}_n$, and $q_n$ are found solving equations~(\ref{eq:lhs41}) to~(\ref{eq:lhs44}).
These equations appear equivalent to an Eulerian multi-fluid model but the source terms ($a_n$, $b_n$, ${\bm c}_n$, and $d_n$) are determined through moment transformation of equation~(\ref{eq:williams}).
Using expressions~(\ref{eq:lhs41}) to~(\ref{eq:lhs44}), this moment transformation of the left hand side of equation~(\ref{eq:williams}) yields the linear system
\begin{align}
\label{eq:lhs9}
\int r^k u_1^l u_2^m u_3^p q^s &\left( \dfrac{\partial f}{\partial t} + \dfrac{\partial {\bm u} f}{\partial {\bm x}} \right) dr du_1 du_2 du_3 dq =
\nonumber \\
& \sum^N_{n=1} (1-k) r^k_n u^l_{1,n} u^m_{2,n} u^p_{3,n} q^s_n a_n
\nonumber \\
+ & \sum^N_{n=1} (-k+l+m+p) r^{k-1}_n u^l_{1,n} u^m_{2,n} u^p_{3,n} q^s_n b_n
\nonumber \\
+ & \sum^N_{n=1} \left( l u^{-1}_{1,n} c_{1,n} + m u^{-1}_{2,n} c_{2,n} + p u^{-1}_{3,n} c_{3,n} \right) r^{k-1}_n u^l_{1,n} u^m_{2,n} u^p_{3,n} q^s_n
\nonumber \\
+ & \sum^N_{n=1} -s r^{k-1}_n u^{l-1}_{1,n} u^{m-1}_{2,n} u^{p-1}_{3,n} q^s_n {\bm c}_n \nonumber \\
+ & \sum^N_{n=1} -s r^{k-1}_n u^{l-1}_{1,n} u^{m-1}_{2,n} u^{p-1}_{3,n} q^{s-1}_n d_n \, .
\end{align}
The complete mathematical derivation is given in the Appendix.

\subsection{Phase-space transport}
\label{sec:rhs}

\textcolor{black}{
The momentum transformation of the right hand side of equation~(\ref{eq:williams}) yields the linear system
\begin{align}
\label{eq:source}
\int r^k \, u_1^l \, u_2^m \, u_3^p \, q^s &\left( - \bm{F}_f - q_f \right) dr \, du_1 \, du_2 \, du_3 \, dq \, =
\nonumber \\
&\sum^N_{n=1} w_n \, r^k_n \, u^{l-1}_{1,n} \, u^m_{2,n} \, u^p_{3,n} \, q^s_n \left(l \dfrac{\partial u_{1,n}}{\partial t} \right)
\nonumber \\
+&\sum^N_{n=1} w_n \, r^k_n \, u^l_{1,n} \, u^{m-1}_{2,n} \, u^p_{3,n} \, q^s_n \left(m \dfrac{\partial u_{2,n}}{\partial t} \right)
\nonumber \\
+&\sum^N_{n=1} w_n \, r^k_n \, u^l_{1,n} \, u^m_{2,n} \, u^{p-1}_{3,n} \, q^s_n \left(p \dfrac{\partial u_{3,n}}{\partial t} \right)
\nonumber \\
+&\sum^N_{n=1} w_n \, r^k_n \, u^l_{1,n} \, u^m_{2,n} \, u^p_{3,n} \, q^{s-1}_n
\left( s \left( \dfrac{\partial q_n}{\partial t} + \nu_q \dfrac{\partial^2 q_n}{\partial {\bm x}^2} \right) \right) \, .
\end{align}
The complete DQMOM linear system is obtained through the quadrature-based approximation and moment transformation of equation~(\ref{eq:williams}).
After these operations, the left-hand side of equation~(\ref{eq:williams}) takes the form of equation~(\ref{eq:lhs9}) and the right-hand side the form of equation~(\ref{eq:source}).
The closure of this system requires the modeling of the terms $\dot{\bm u}$ and $\dot{q}$, which is described in the following section.
The exact form of the DQMOM linear system depends on the moments ($k$, $l$, $m$, $p$, $s$) which are chosen avoiding a singular coefficient matrix.
After determining the source terms, the DQMOM equation system comprises $6 \times N$ unknowns, namely $a_n$, $b_n$, ${\bm c}_n$, and $d_n$.
In case of a two-dimensional simulation the number of unknowns reduces to $5 \times N$.
To set up a well-defined linear equation system, we need choose a number of combinations of moments which equals the number of unknowns. 
Thus, our two-dimensional simulations and $N=3$ require 15 moment combinations.
Our choice of moments is given by the columns of the matrix
\begin{equation}
\label{eq:moments}
\begin{Bmatrix} 
k \\ l \\ m \\ p \\ s
\end{Bmatrix} 
=
\setcounter{MaxMatrixCols}{20}
\begin{Bmatrix} 
3 & 3 & 2 & 1 & 2 & 0 & 1 & 0 & 0 & 3 & 2 & 1 & 1 & 1 & 1 \\
0 & 1 & 1 & 0 & 0 & 0 & 1 & 1 & 1 & 0 & 0 & 0 & 1 & 1 & 1 \\
0 & 0 & 0 & 0 & 0 & 1 & 0 & 1 & 0 & 1 & 1 & 1 & 1 & 1 & 1 \\
0 & 0 & 0 & 0 & 0 & 0 & 0 & 0 & 0 & 0 & 0 & 0 & 1 & 1 & 1 \\
0 & 0 & 0 & 0 & 0 & 0 & 0 & 0 & 0 & 0 & 0 & 0 & 1 & 2 & 3
\end{Bmatrix} 
\, .
\end{equation}
}

\subsection{Source terms}
\label{sec:source}

The closure of equations~(\ref{eq:williams}), (\ref{eq:lhs9}), and~(\ref{eq:source}) requires the modeling of the source terms, namely $\dot{\bm u}_n$ and $\dot{q}_n$.
In the following, the equations are given for discrete particles, thus, the indices to the nodes $n$ are omitted.
The acceleration of a particle is affected by the aerodynamic drag forces and the electrostatic forces given by
\begin{equation}
\label{eq:udot}
\dot{\bm u} = 
\frac{3}{8} \frac{ \rho_{\rm g}}{r \, \rho} ({\bm u}_\mathrm{g} - {\bm u}) |{\bm u}_\mathrm{g} - {\bm u}| C_\mathrm{d}
\textcolor{black}{\,+\, \dfrac{3}{4} \, \dfrac{q \, {\bm E}}{\pi r^3 \rho}} \, ,
\end{equation}
where $\rho_\mathrm{g}$ and ${\bm u}_\mathrm{g}$ are the density and velocity of the surrounding gas and $\rho$ is the material density of the particle.
The drag coefficient, $C_\mathrm{d}$, is determined as\citep{Put61}
\begin{equation}
\label{eq:cd}
C_\mathrm{d} = 
\left\{ 
\begin{matrix}
\frac{24}{Re_\mathrm{p}}\left(1 +0.15\ Re_\mathrm{p}^{0.687}\right) & \quad \mbox{for \ \ \ }Re_\mathrm{p} < 10^3 \\
0.44 & \quad \mbox{\ for \ \ \ }Re_\mathrm{p} \ge 10^3 \, .
\end{matrix}
\right.
\end{equation}
In this equation, the particle Reynolds number is $Re_\mathrm{p} = 2 \,r |{\bm u}_{\rm g} - {\bm u}| / \nu_\mathrm{g}$ with $\nu_\mathrm{g}$ being the kinematic viscosity of the gas.

\textcolor{black}{
The last term in equation~(\ref{eq:udot}) describes the electrostatic force acting on a given particle in case it is subjected to an external electrostatic field.
As mentioned above, the electric field strength ${\bm E}$ is computed by Gauss's law.
In the electrostatic approximation,  ${\bm E}$ is defined in terms of the electric potential $\varphi$,
\begin{equation}
\label{eq:elecpot}
{\bm E} \;=\; -\nabla \varphi\,,
\end{equation}
so that Gauss's law reduces to,
\begin{equation}
\label{eq:gauss}
\nabla^2 \varphi \;=\; -\dfrac{\rho_{\mathrm{el}}}{\varepsilon} \,,
\end{equation}
which is a Poisson equation.
In this equation, ${\rho_{\mathrm{el}}}$ is the electric charge density, which is numerically approximated by the ratio of the total charge of all particles located in one cell and the volume of the cell.
Since no external electric field is considered, the electric charge density results directly from the location and the charge of the surrounding particles.
The solid volume fraction in our simulations is small, thus, we apply a value of $\varepsilon=8.85 \times 10^{-12}$~F/m for the electric permittivity of the solid-gaseous mixture.
}

As discussed above, $\dot{q}$ reflects the accumulation of charge upon contact with surfaces through the triboelectric effect.
However, we assume that the particles close to the walls often collide with the walls and, thus, reach their equilibrium charge.
Thus, the effect of charge uptake during particle-wall collisions is not included in the term $\dot{q}$ but instead in the boundary conditions imposed to the field of $q$.
Further, the charge transfer between particles during particle-particle collisions which leads to the spatial redistribution of charge is included in the charge diffusion term, which means the last term of equation~(\ref{eq:qf}).

The local effect of inter-particle collisions, on the other hand, need to be handled explicitly by modeling $\dot{q}$.
In other words, $\dot{q}$ accounts for the charge transfer between particles located within the same infinitesimal volume, for example in a computational cell.
In the present study, we consider the powder to be composed of homogeneous material and, therefore, no charge exchange due to a contact potential difference is assumed between equally charged particles.
However, for colliding particles carrying different charges, a charge transfer mechanism is modeled using the analogy to a capacitor as proposed by \citet{soo1971dynamics}. 
According to this analogy, the exchanged charge between two particles of the initial charges $q_1$ and $q_2$ and capacities of $C_1$ and $C_2$ is\citep{soo1971dynamics}
\begin{equation}
\Delta q_{1} = - \Delta q_{2} = \dfrac{C_{1} C_{2}}{C_{1} + C_{2}} \left (\dfrac{q_{2}}{C_{2}} - \dfrac{q_{1}}{C_{1}} \right) \left( 1 - e^{-\Delta t_{12} / T_{12}} \right) \, .
\end{equation}
In the above equation, the contact time of the collision is $\Delta t_{12} = 2.94 \, \alpha \left | \bm{u}_{p,12} \right|$.
The charge relaxation time is 
\begin{equation}
T_{12} = \dfrac{C_{1} C_{2}}{C_{1} + C_{2}} \dfrac{r_1 + r_2}{A_{12}} \varphi_{p} \, ,
\end{equation}
where the electric capacity of a spherical particle is given by $C = 4 \pi \epsilon _{0} r$ and $\varphi_{p}$ denotes the resistivity of the particle.
The contact surface $A_{12}$ is expressed as
\begin{equation}
A_{12} = \dfrac{\pi r_1 r_2}{r_1 + r_2} \alpha \, .
\end{equation}
In the above equations, $\alpha$ is following the elastic theory of Hertz\citep{soo1971dynamics}
\begin{equation}
\alpha = r_1 r_2 \left (\dfrac{5}{4}\pi \rho \left |\bm{u}_{12} \right |^2 \dfrac{\sqrt{r_1} + r_2}{r^3_1 + r^3_2} \dfrac{1 - \mu^2}{E}\right )^{2/5} \, ,
\end{equation}
with $\bm{u}_{12}$ being the relative velocity between the colliding particles and $\mu$, and $E$ is the Poisson ratio and the Young modulus of the particulate material, respectively.

Besides the charge also the velocity of the particles is updated assuming fully elastic collisions.
It is noted that, contrary to our earlier Eulerian-Lagrangian simulations,\citep{Gro17e} electrostatic forces acting on the particles during their interaction are not considered.
The frequency of binary particle-particle collisions is determined through a stochastic approach based on a Poisson probability distribution function.\cite{Gro19d}
The assumption of binary particle collisions is justified for dilute flows.

Substitution of these models for $\dot{\bm u}$ and $\dot{q}$ into equation~(\ref{eq:source}) results in a linear system of equations, and its solution provides the source terms required in equations~(\ref{eq:lhs41}) to~(\ref{eq:lhs44}), which then constitutes a closed system.

\subsection{Numerical solution strategy}

\textcolor{black}{
The finite difference method is used to find a numerical solution to the equation system.
The first-order derivatives in equations~(\ref{eq:lhs41}) to~(\ref{eq:lhs44}) and (\ref{eq:source}) are approximated by a first-order upwind scheme in downstream direction and by second-order central schemes in wall-normal direction.
Second-order central schemes are also used to discretize the second-order derivatives in equation~(\ref{eq:source}) and on the left-hand side of Gauss's law \eqref{eq:gauss}.
The discretized equations are solved iteratively using the Jacobi algorithm.
The iterations proceed until the $L_2$-error norm, which is the rms of the difference between two subsequent iterations, of the particle charge drops by five orders of magnitude.
}

\section{Simulation conditions}
\label{numset}

We simulated eight generic flow cases to test the newly developed numerical framework.
More specifically, we considered a two-dimensional steady particle-laden channel flow.
The particles occupy a volume of 1~\%, which means the flow is dilute.
The chosen setup aims to provide insights in the behavior of the new method and its response to parameter variations.
Thus, we studied idealized conditions rather than realistic flows.

The numerical domain is bounded by two parallel planar walls, an inlet, and an outlet.
According to our coordinate convention, the streamwise direction is denoted by the $x$-axis and the wall-normal direction by the $y$-axis. 
The width of the channel in the $y$-direction is $H=$~40~mm and its length in the $x$-direction is 10\,$H$.
Boundary conditions are applied to the variables of the particulate field at the sidewalls, the inlet, and the outlet as detailed below.

\begin{figure}[b]
\centering
\includegraphics[trim=0cm 0cm 0cm 0cm,clip=true,width=0.47\textwidth]{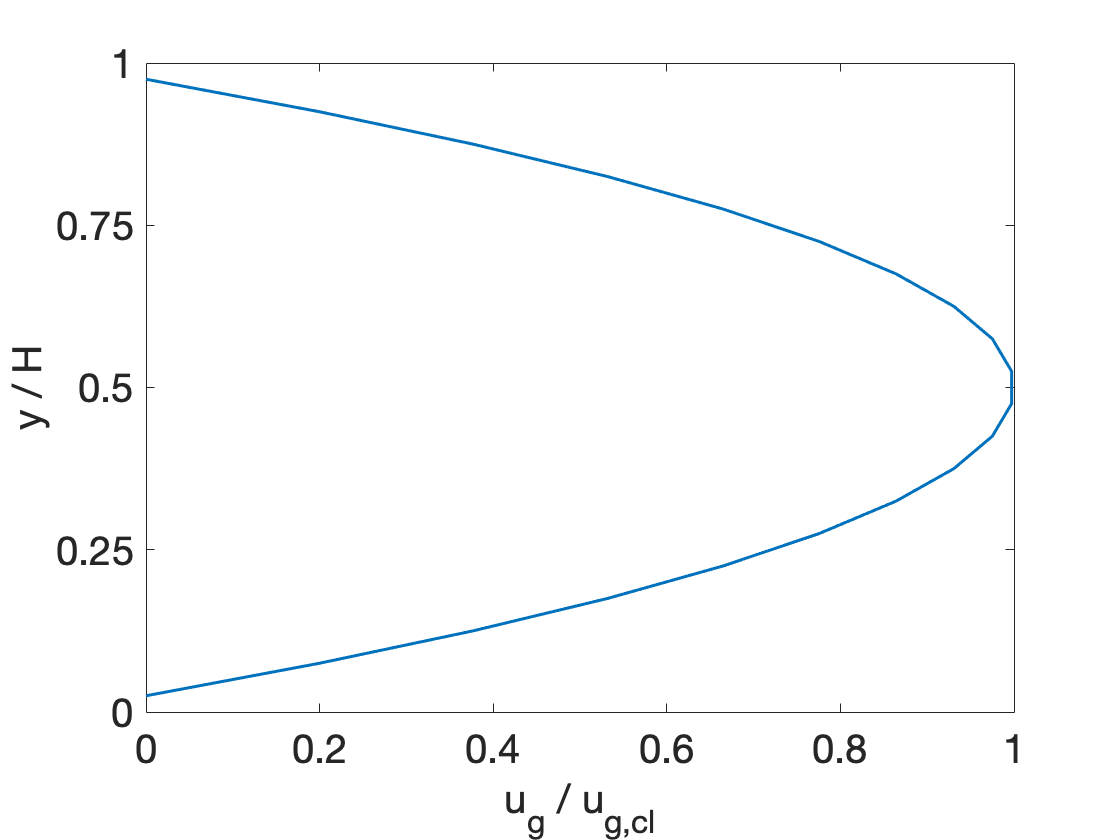}
\caption{Constant profile of the streamwise gaseous velocity.}
\label{fig:medium_gas}
\end{figure}
As regards the gaseous phase, we assumed a steady and in streamwise direction constant velocity profile, $u_\mathrm{g}(y)$, as depicted in Figure~\ref{fig:medium_gas}.
According to this profile, no-slip boundary conditions are applied at the side walls and the maximum streamwise velocity at the centreline of the channel is $u_\mathrm{g,cl}=5$\,m/s.
Further, we used a zero wall-normal gas velocity.
The density $\rho_\mathrm{g}=1.225$~kg/m$^3$ and kinematic viscosity $\nu_\mathrm{g}=1.48\times10^{-5}$~m$^2$/s of the gas are constant.
Thus, the Reynolds number based on the channel width and the average gas velocity is about 10\,000.
The condition of the gaseous phase is the same for all simulated cases.

We assume the walls of the channel to be conductive and grounded.
Thus, the electric potential is zero at the walls and we apply a zero-gradient boundary condition at the in- and outlet.

\renewcommand{\arraystretch}{0.8}
\begin{table*}[tb]
\caption{Nodes of the approximations to the inlet ($x=0$) and wall ($y=0$ and $y=H$) particle distribution functions and parameters of the simulated test cases.}
\label{tab:table1}
\begin{ruledtabular}
\begin{tabular}{cl|cccccc}
Case  & $n$ \hspace{5mm} & $w_{n,x=0}$ & $r_{n,x=0}$ & $u_{1,n,x=0}$ & $q_{n,x=0}$ & $q_{\mathrm{eq},n}$ &$\nu_q$ \\
      &                  & $w_{n,y=0/H}$ & $r_{n,y=0/H}$ & $u_{1,n,y=0/H}$ & $q_{n,y=0/H}$ &                     &        \\
& & (-) & ($\upmu$m)  & (m/s) & ($\upmu$C) & ($\upmu$C) & (mm$^2$/s) \\ \hline
      & 1  & 0.5 & 25  & $u_\mathrm{g}(y) + 1.02$ & 0.01 & 1.1 & 100 \\
A     & 2  & 0.3 & 50  & $u_\mathrm{g}(y) + 1.03$ & 0.02 & 1.2 & 100 \\
      & 3  & 0.2 & 150 & $u_\mathrm{g}(y) + 1.05$ & 0.03 & 1.3 & 100 \\
\hline                                                    
      & 1  & 0.5 & 25  & $u_\mathrm{g}(y) + 2.02$ & 0.01 & 1.1 & 100 \\
B     & 2  & 0.3 & 50  & $u_\mathrm{g}(y) + 2.03$ & 0.02 & 1.2 & 100 \\
      & 3  & 0.2 & 150 & $u_\mathrm{g}(y) + 2.05$ & 0.03 & 1.3 & 100 \\
\hline                                                    
      & 1  & 0.5 & 75  & $u_\mathrm{g}(y) + 1.02$ & 0.01 & 1.1 & 100 \\
C     & 2  & 0.3 & 100 & $u_\mathrm{g}(y) + 1.03$ & 0.02 & 1.2 & 100 \\
      & 3  & 0.2 & 200 & $u_\mathrm{g}(y) + 1.05$ & 0.03 & 1.3 & 100 \\
\hline                                                    
      & 1  & 0.5 & 25  & $u_\mathrm{g}(y) + 1.02$ & 0.01 & 1.1 & 20 \\
D     & 2  & 0.3 & 50  & $u_\mathrm{g}(y) + 1.03$ & 0.02 & 1.2 & 20 \\
      & 3  & 0.2 & 150 & $u_\mathrm{g}(y) + 1.05$ & 0.03 & 1.3 & 20 \\
\hline                                                    
      & 1  & 0.5 & 75  & $u_\mathrm{g}(y) + 2.02$ & 0.01 & 1.1 & 100 \\
E     & 2  & 0.3 & 100 & $u_\mathrm{g}(y) + 2.03$ & 0.02 & 1.2 & 100 \\
      & 3  & 0.2 & 200 & $u_\mathrm{g}(y) + 2.05$ & 0.03 & 1.3 & 100 \\
\hline                                                    
      & 1  & 0.5 & 75  & $u_\mathrm{g}(y) + 1.02$ & 0.01 & 1.1 & 20 \\
F     & 2  & 0.3 & 100 & $u_\mathrm{g}(y) + 1.03$ & 0.02 & 1.2 & 20 \\
      & 3  & 0.2 & 200 & $u_\mathrm{g}(y) + 1.05$ & 0.03 & 1.3 & 20 \\
\hline                                                    
      & 1  & 0.5 & 25  & $u_\mathrm{g}(y) + 2.02$ & 0.01 & 1.1 & 20 \\
G     & 2  & 0.3 & 50  & $u_\mathrm{g}(y) + 2.03$ & 0.02 & 1.2 & 20 \\
      & 3  & 0.2 & 150 & $u_\mathrm{g}(y) + 2.05$ & 0.03 & 1.3 & 20 \\
\hline                                                    
      & 1  & 0.5 & 75  & $u_\mathrm{g}(y) + 2.02$ & 0.01 & 1.1 & 20 \\
H     & 2  & 0.3 & 100 & $u_\mathrm{g}(y) + 2.03$ & 0.02 & 1.2 & 20 \\
      & 3  & 0.2 & 200 & $u_\mathrm{g}(y) + 2.05$ & 0.03 & 1.3 & 20 \\
\end{tabular}
\end{ruledtabular}
\end{table*}
\renewcommand{\arraystretch}{1.0}

The inlet and wall boundary conditions and properties of the particulate phase are varied.
At the outlet, we applied a zero-gradient boundary condition to all variables.
More specifically, we varied three parameters, the initial radii and streamwise velocity distribution of the particles and the charge diffusion coefficient.
In total, these choices result in eight simulation cases, denoted by the letters A to G, as given in Table~\ref{tab:table1}.
It is reiterated that continuous particle size, velocity, and charge distributions are considered.
As mentioned above, we applied a quadrature approximation of $N=3$ nodes to these distributions.
Thus, $f$ is approximated by three nodes, each of which is characterized by an initial weight, radius, streamwise ($u_1$) and wall-normal ($u_2$) velocity, and charge, as documented in Table~\ref{tab:table1}.

The distribution of the initial weights on the three nodes ($w_{n,x=0}$) is the same for each case and their sum yields unity.
Two initial radius distributions are considered:
one representing smaller particle sizes ($r_{n,x=0}=25\,\upmu$m, $50\,\upmu$m, and $150\,\upmu$m) and one larger ones ($r_{n,x=0}=75\,\upmu$m, $100\,\upmu$m, and $200\,\upmu$m).
The initial streamwise velocity distribution varies in wall-normal direction with a positive offset to the velocity of the gas phase.
Two different offsets represent initially faster and slower particles.
As can be seen in Table~\ref{tab:table1}, the assigned offset is slightly different for each node, for example, 1.02\,m/s, 1.03\,m/s, and 1.05\,m/s for case~A.
By doing so, singularities in the solution of the linear equation system, which would be caused if different nodes adopt at one location identical numerical values, are avoided.
Similarly, the initial wall-normal particle velocities are chosen to be non-zero and slightly different from each other to avoid singularities.
However, the values of $u_{2,n,x=0}$ and $u_{2,n,y=0/H}$ are of the order $10^{-10}$ to resemble a nearly zero net wall-normal velocity.

Further, the particles initially carry only little charge.
As mentioned before, we assume that the particles which are located close to the walls, respectively in the computational cells adjacent to the walls, obtain their equilibrium charge due to frequent wall contacts.
It is emphasized that this assumption is not contradictory to the small values for $u_{1,n}$.
Instead, $u_{1,n}$ is an Eulerian field variable which represents by definition the average over all particles present in one cell.
Thus, each particle can have a non-zero wall-normal velocity.
The values of the size-dependent equilibrium charges given in Table~\ref{tab:table1} are approximated from the data of \citet{Mat18}. 
Our study aims to test the response of our model to individual parameter variations.
Therefore, to avoid the simultaneous variation of two parameters, we decided to keep the equilibrium charge distribution even though the size distribution is changed (compare for example cases~A and~C).

Finally, all particles are of a material density of 1225~kg/m$^3$ or a thousand times the gaseous density.
Moreover, the solid to fluid volume ratio is for all simulations set to 1\%.

\section{Results and discussion}
\label{results}

\subsection{Grid convergence}

\begin{figure}[b]
\centering
\subfigure[]{\includegraphics[trim=0cm 0cm 0cm 0cm,clip=true,width=0.47\textwidth]{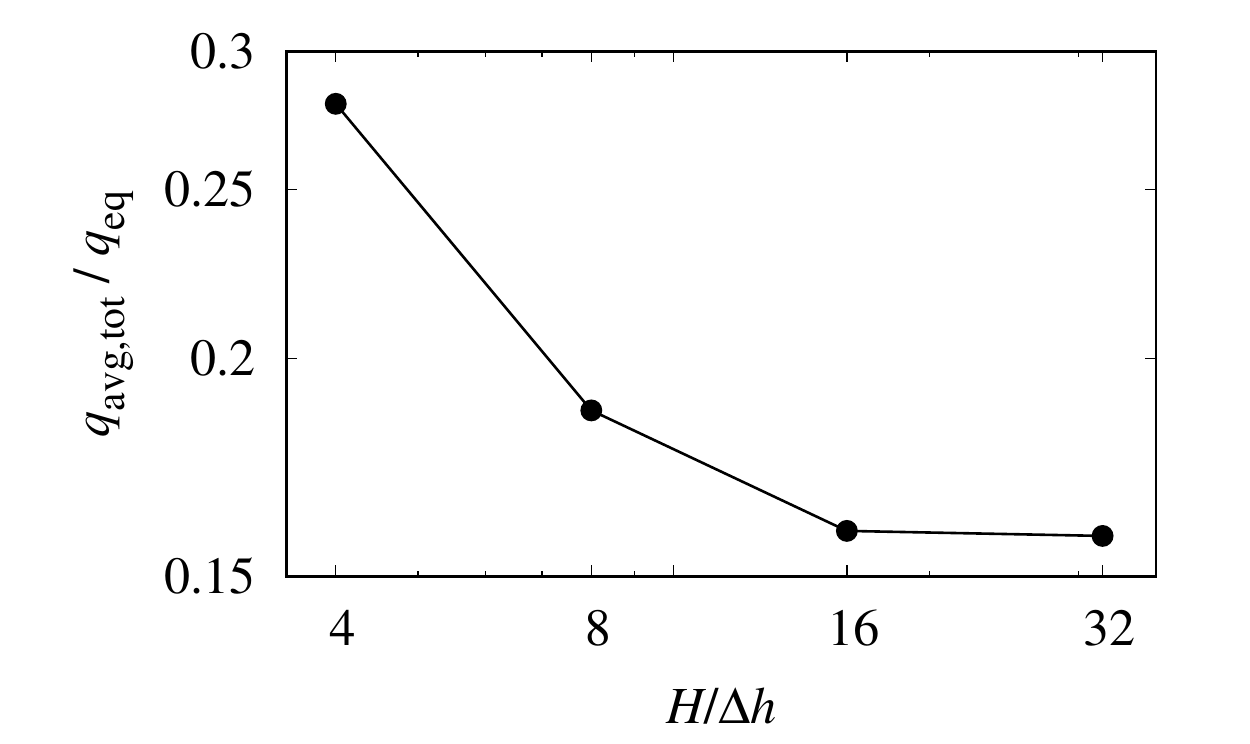}\label{fig:2a}}
\quad
\subfigure[]{\includegraphics[trim=0cm 2.0cm 0cm 5.5cm,clip=true,width=0.47\textwidth]{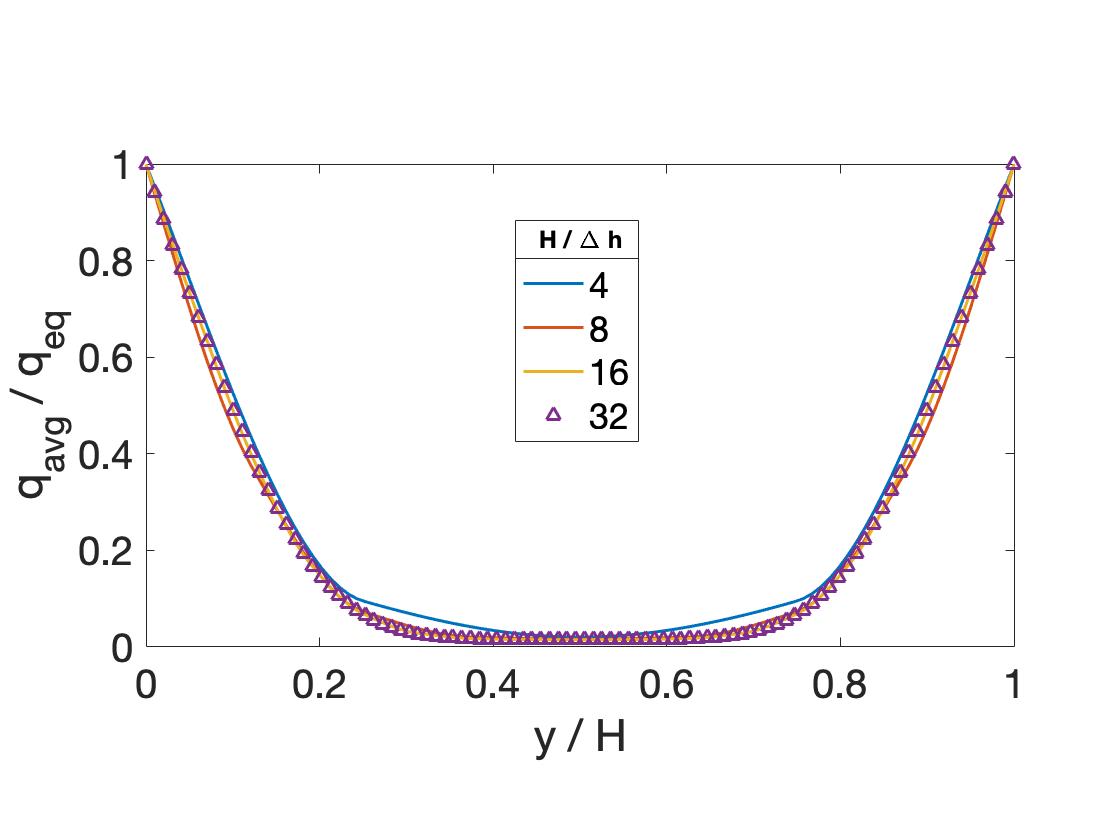}\label{fig:2b}}
\caption{Convergence of the average charge of (a) all particles and (b) at the outlet cross-section ($x=10\,H$) for increasing grid resolution.}
\label{fig:convergence}
\end{figure}

First, the optimal grid resolution for the numerical solution of the governing equations is determined.
For simplicity, the mesh is Cartesian, uniform, and rectilinear.
For the evaluation of the convergence behavior, we simulated case A (cf.~Table~\ref{tab:table1}) on four different meshes varying between $4\times40=160$ and $32\times320=10\,240$ cells.
This simulation and all cases in this paper regard a two-dimensional steady channel flow.
The average charge of all particles present in the system, which is given by 
\begin{equation}
q_\mathrm{avg,tot} = \frac{\sum\limits_{n,i,j} q_{n,i,j} \,  w_{n,i,j}}{\sum\limits_{n,i,j} w_{n,i,j}} \, ,
\end{equation}
where $i$ and $j$ are the indices of the cells, is plotted in Figure~\ref{fig:2a}.
Here, the resolution is given in terms of the number of cells over the channel with, $H/\Delta h$.
In this and the following figures the charge is normalized by the average equilibrium charge of the powder at the inlet,
\begin{equation}
\label{eq:qeq}
q_\mathrm{eq} = \frac{\sum\limits_{n} q_{\mathrm{eq},n} \,  w_{n,x=0}}{\sum\limits_{n} w_{n,x=0}} \, ,
\end{equation}
which is according to Table~\ref{tab:table1} for all cases $q_\mathrm{eq}=1.17\,\upmu$C.
Moreover, Figure~\ref{fig:2b} depicts the profile of the average particle charge at the outlet cross-section for each grid resolution.

\textcolor{black}{
As can be seen in the figure, the simulations converged sufficiently for a grid consisting of $16\times160=2560$ cells.
The charge distribution at the outlet collapses for the resolutions of $H/\Delta h=$~16 and~32.
All the results presented in the remainder of this section are computed using a resolution of $32\times320=10\,240$ cells.
Based on the total average charge computed with the cell sizes of $H/\Delta h=$~4, 8, and 16, the overall order-of-accuracy\citep{Roa97} of the numerical method is estimated to~1.76.}

\subsection{Simulation of case A}

\begin{figure}[b]
\begin{center}
\subfigure[]{
\begin{tikzpicture}
\begin{scope}[x=10,y=10]
	\draw [->] (0,0) -- (0,2) node[anchor=west]{\small $y$};         
	\draw [->] (0,0) -- (1.5,0) node[anchor=south]{\small $x$};         
\end{scope}
\end{tikzpicture}
\includegraphics[trim=1cm 14cm 1cm 14cm,clip=true,width=0.82\textwidth]{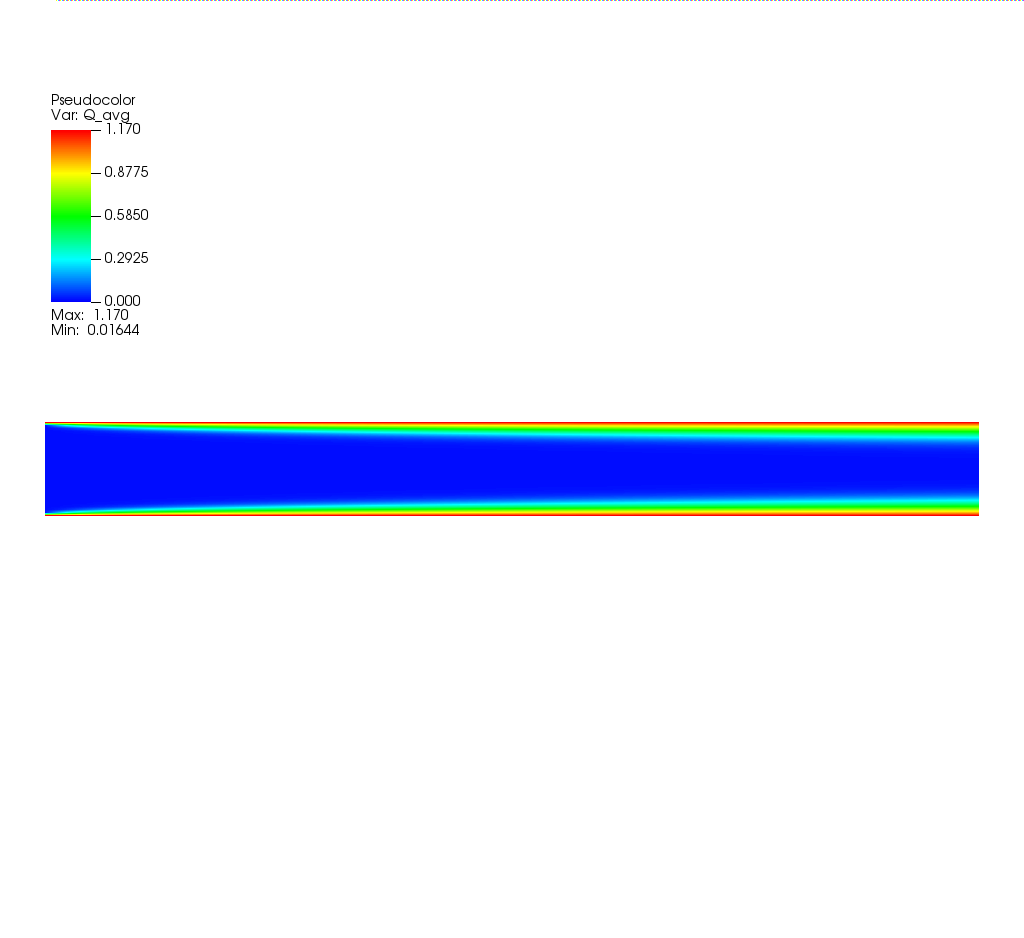}\label{fig:q}}
\begin{tikzpicture}
    \node[anchor=south west,inner sep=0] (Bild) at (0,0)
    {\includegraphics[trim=0cm 0cm 0cm 0cm,clip=true,width=0.025\textwidth,height=0.09\textwidth]{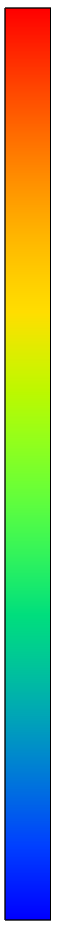}};
    \begin{scope}[x=(Bild.south east),y=(Bild.north west)]
	\draw [](0.9,1.0) node[anchor=south]{$q_\mathrm{avg}/q_\mathrm{eq}$};         
	\draw [](0.7,0.90) node[anchor=west]{\small 1};         
	\draw [](0.7,0.02) node[anchor=west]{\small 0};         
    \end{scope}
\end{tikzpicture}\\
\subfigure[]{
\begin{tikzpicture}
    \begin{scope}[x=10,y=10]
	\draw [->] (0,0) -- (0,2) node[anchor=west]{\small $y$};         
	\draw [->] (0,0) -- (1.5,0) node[anchor=south]{\small $x$};         
    \end{scope}
\end{tikzpicture}
\includegraphics[trim=1cm 14cm 1cm 14cm,clip=true,width=0.82\textwidth]{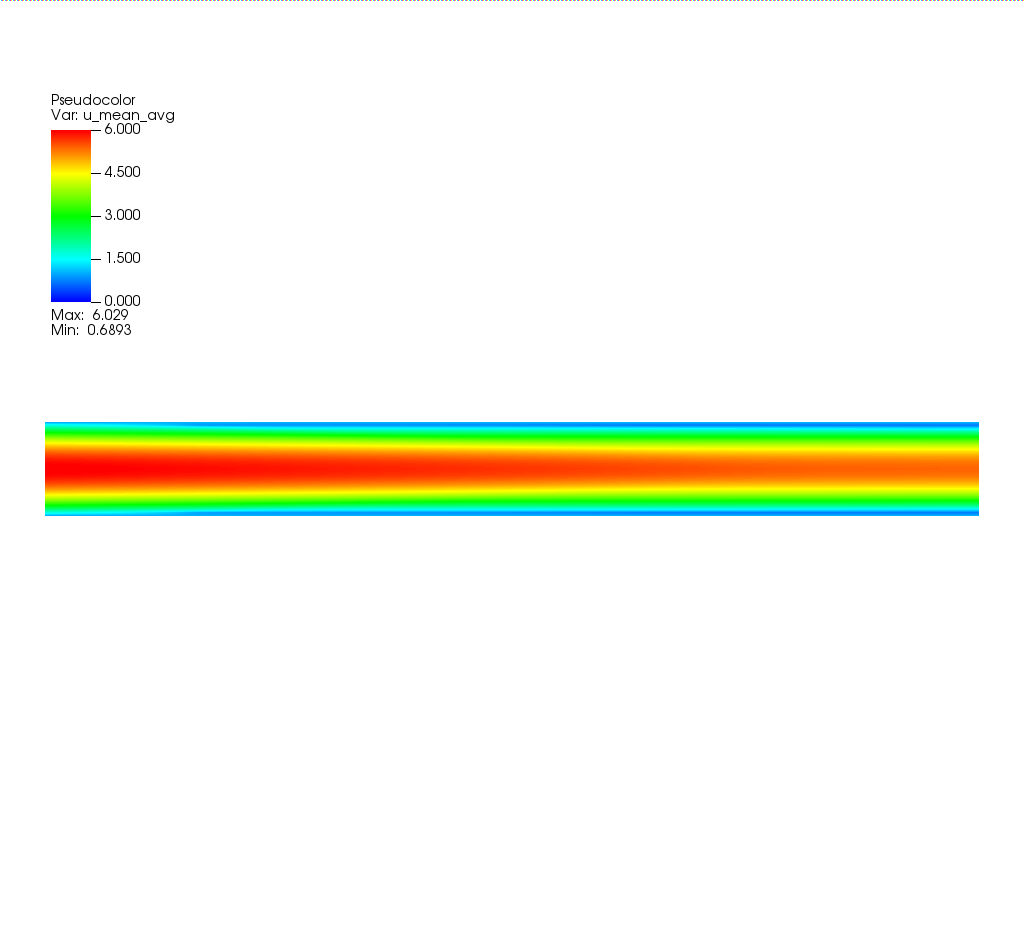}\label{fig:u}}
\hspace{-2mm}
\begin{tikzpicture}
    \node[anchor=south west,inner sep=0] (Bild) at (0,0)
    {\includegraphics[trim=0cm 0cm 0cm 0cm,clip=true,width=0.025\textwidth,height=0.09\textwidth]{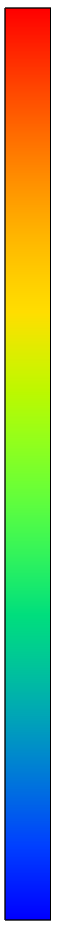}};
    \begin{scope}[x=(Bild.south east),y=(Bild.north west)]
	\draw [](0.9,1.0) node[anchor=south]{$u_\mathrm{avg}/u_\mathrm{g,cl}$};
	\draw [](0.7,0.90) node[anchor=west]{\small 1.2};         
	\draw [](0.7,0.02) node[anchor=west]{\small 0};         
    \end{scope}
\end{tikzpicture}\\
\subfigure[]{
\begin{tikzpicture}
    \begin{scope}[x=10,y=10]
	\draw [->] (0,0) -- (0,2) node[anchor=west]{\small $y$};         
	\draw [->] (0,0) -- (1.5,0) node[anchor=south]{\small $x$};         
    \end{scope}
\end{tikzpicture}
\includegraphics[trim=1cm 14cm 1cm 14cm,clip=true,width=0.82\textwidth]{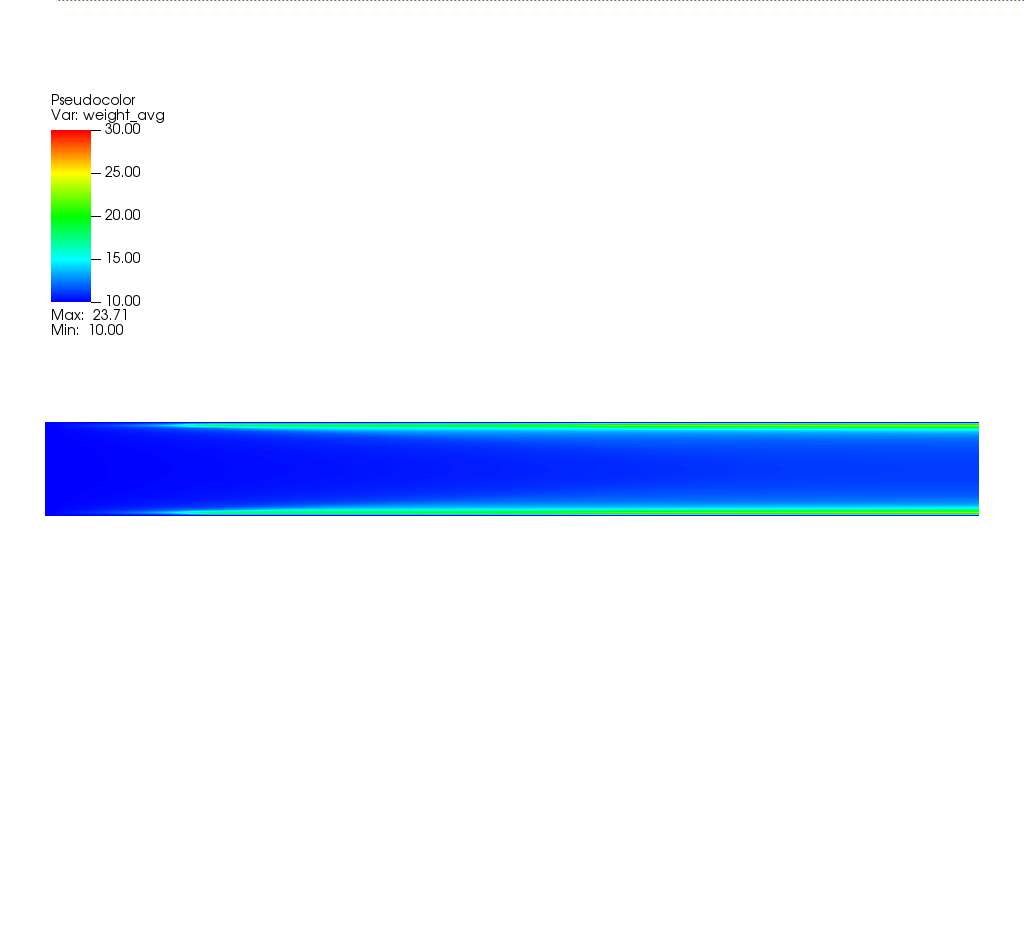}\label{fig:w}}
\begin{tikzpicture}
    \node[anchor=south west,inner sep=0] (Bild) at (0,0)
    {\includegraphics[trim=0cm 0cm 0cm 0cm,clip=true,width=0.025\textwidth,height=0.09\textwidth]{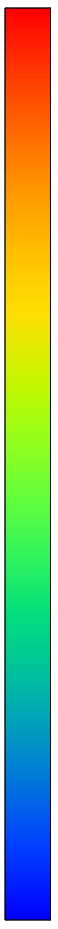}};
    \begin{scope}[x=(Bild.south east),y=(Bild.north west)]
	\draw [](0.9,1.0) node[anchor=south]{$w_\mathrm{sum}/w_\mathrm{in}$};         
	\draw [](0.7,0.90) node[anchor=west]{\small 3};         
	\draw [](0.7,0.02) node[anchor=west]{\small 1};         
    \end{scope}
\end{tikzpicture}\\
\subfigure[]{
\begin{tikzpicture}
    \begin{scope}[x=10,y=10]
	\draw [->] (0,0) -- (0,2) node[anchor=west]{\small $y$};         
	\draw [->] (0,0) -- (1.5,0) node[anchor=south]{\small $x$};         
    \end{scope}
\end{tikzpicture}
\includegraphics[trim=1cm 14cm 1cm 14cm,clip=true,width=0.82\textwidth]{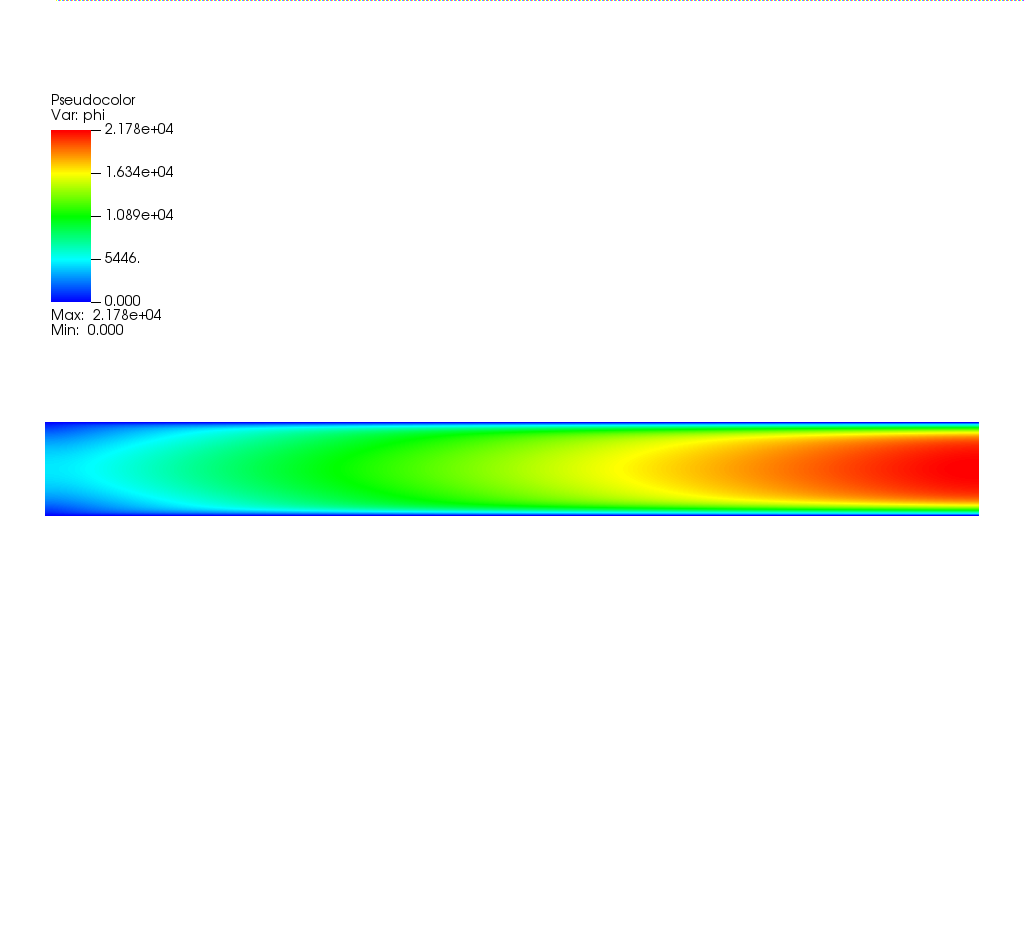}\label{fig:phi}}
\begin{tikzpicture}
    \node[anchor=south west,inner sep=0] (Bild) at (0,0)
    {\includegraphics[trim=0cm 0cm 0cm 0cm,clip=true,width=0.025\textwidth,height=0.09\textwidth]{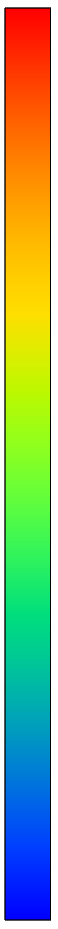}};
    \begin{scope}[x=(Bild.south east),y=(Bild.north west)]
	\draw [](0.9,1.0) node[anchor=south]{$\varphi/\varphi_\mathrm{max}$};         
	\draw [](0.7,0.90) node[anchor=west]{\small 1};         
	\draw [](0.7,0.02) node[anchor=west]{\small 0};         
    \end{scope}
\end{tikzpicture}
\end{center}
\caption{Normalized average (a) particle charge, (b) downstream velocity, (c) total number density, and (d) electric potential for case~A.}
\label{fig:2d}
\end{figure}
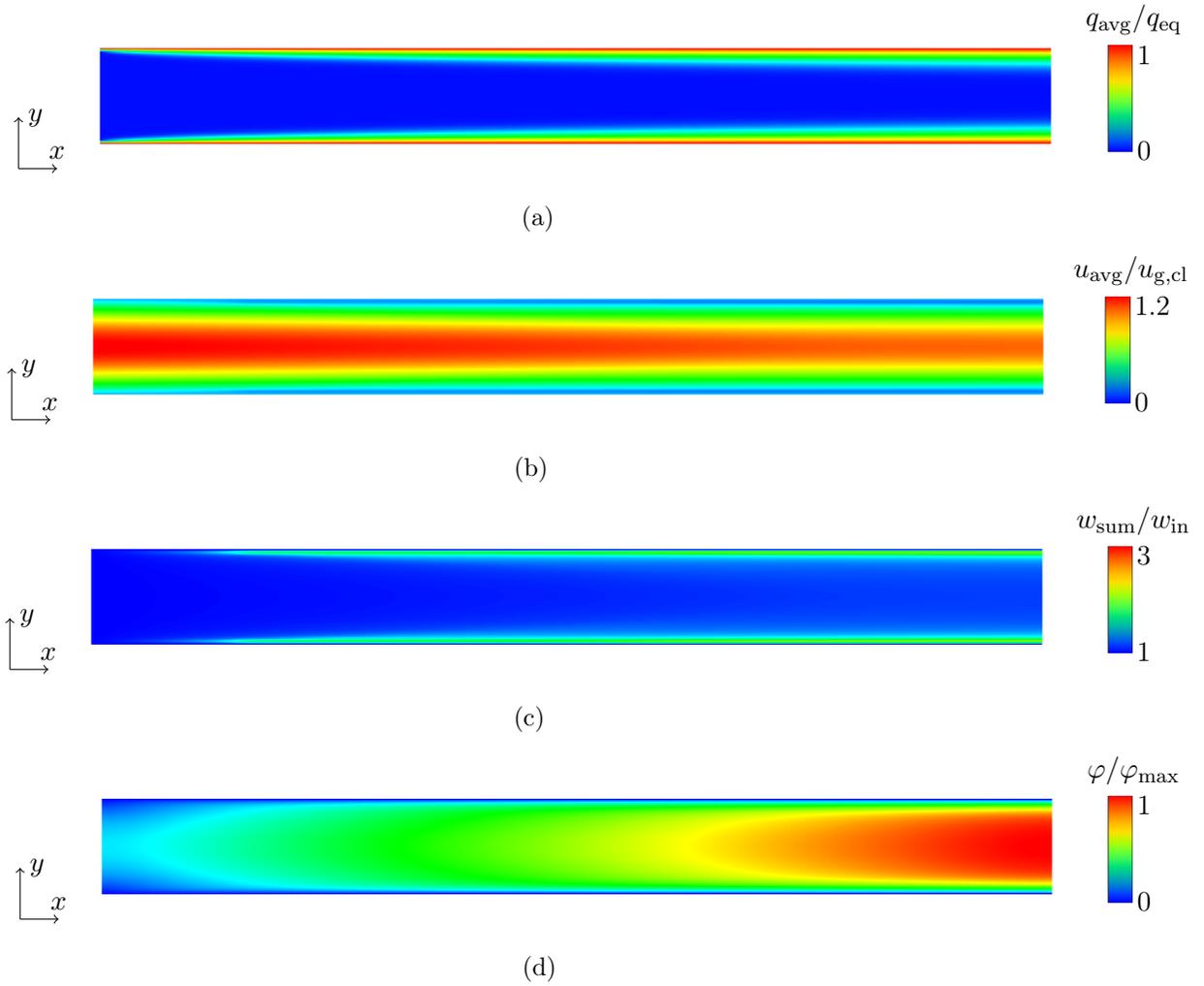

The newly developed solver can predict Eulerian fields of charging polydisperse powder flow.
This is demonstrated by Figure~\ref{fig:2d} which depicts the properties of the particle flow through the channel corresponding to the conditions of case~A.
The first plot, Figure~\ref{fig:q}, shows the local average particle charge.
In the left, the blue color represents the low charged particles entering the domain at the inlet.
The red zone close to the upper and lower boundary consists of those particles which obtained their maximum charge due to collisions with the wall.

The graph correctly reflects the increase of particle charge in the downstream direction.
While flowing in positive $x$-directions more particles gain charge at the walls.
This accumulated charge convects at the same time downstream and migrates towards the center of the channel through diffusion.
Due to this motion in both spatial directions, the triangular shape of the low charged particles, i.e., the blue area at the inlet in Figure~\ref{fig:q}, forms.
Also, this motion is responsible for the growth of the red area towards the outlet which represents the transport of charge towards the particles in the wall-normal direction.
This can be interpreted as a charge boundary layer.
According to the discussion, the thickness of this charge boundary layer depends on the balance of the charge motion perpendicular and parallel to the wall.
As such, the charge boundary layer develops analogously to the momentum, species, thermal, or other types of boundary layers.

Further, the evolution of the streamwise particle velocity is given in Figure~\ref{fig:u}.
The fact that the particles move faster in the center of the channel directly relates to the velocity profile imposed as inlet condition.
The particles slow down in the downstream direction which results from the drag forces of the gas.
As noted above, the wall-normal velocity of the particles is very small throughout the domain and, thus, omitted here.

Next, the particle weights or number densities are plotted in Figure~\ref{fig:w}.
As expected, the relative number of particles grows in the regions of low streamwise velocity, namely close to the walls.
Finally, Figure~\ref{fig:phi} shows the electric potential which arises from the charge carried by the particles.
Since the walls are conductive, $\varphi$ becomes zero at the upper and lower boundary.
Moreover, the electric potential increases in streamwise direction following the increase of the charge of the particles.

Overall, the new solver predicts the charging behavior in an expected way.

\subsection{Variation of individual simulation parameters}

\begin{figure}[p]
\begin{center}
\subfigure[Case A.]{
\includegraphics[trim=0cm 2cm 0cm 4cm,clip=true,width=0.44\textwidth]{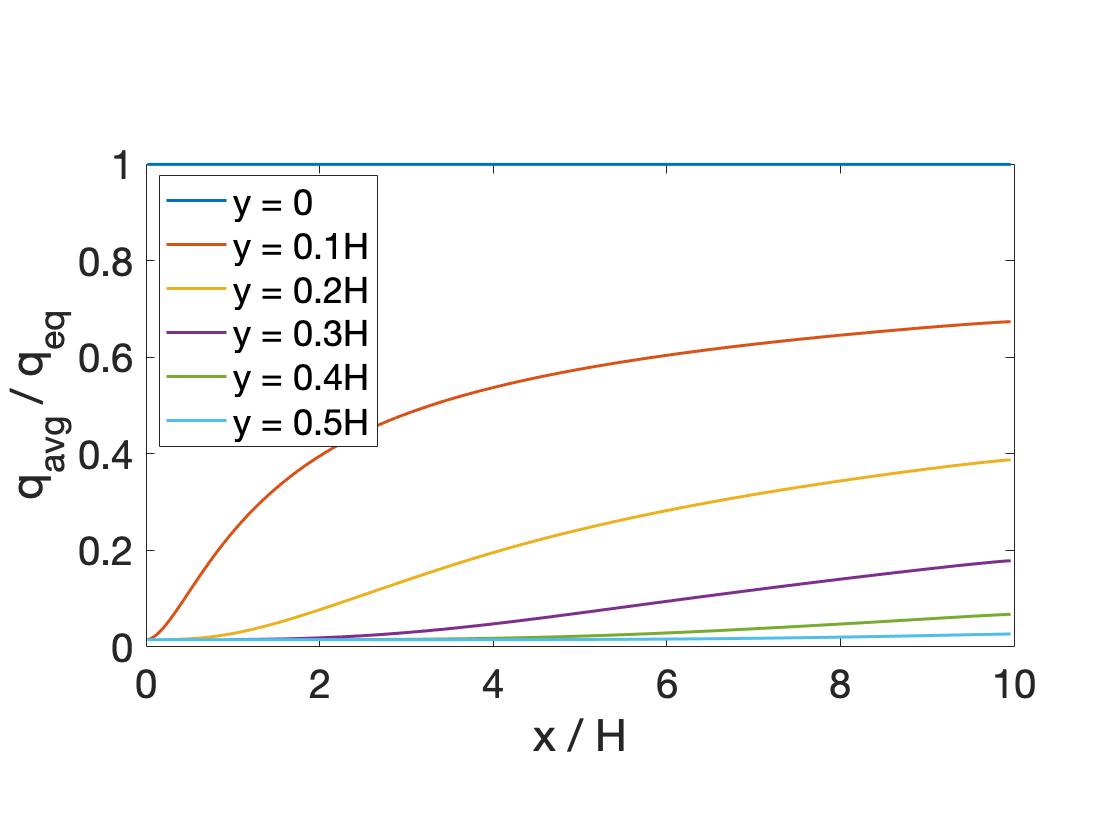}\qquad
\includegraphics[trim=0cm 2cm 0cm 4cm,clip=true,width=0.44\textwidth]{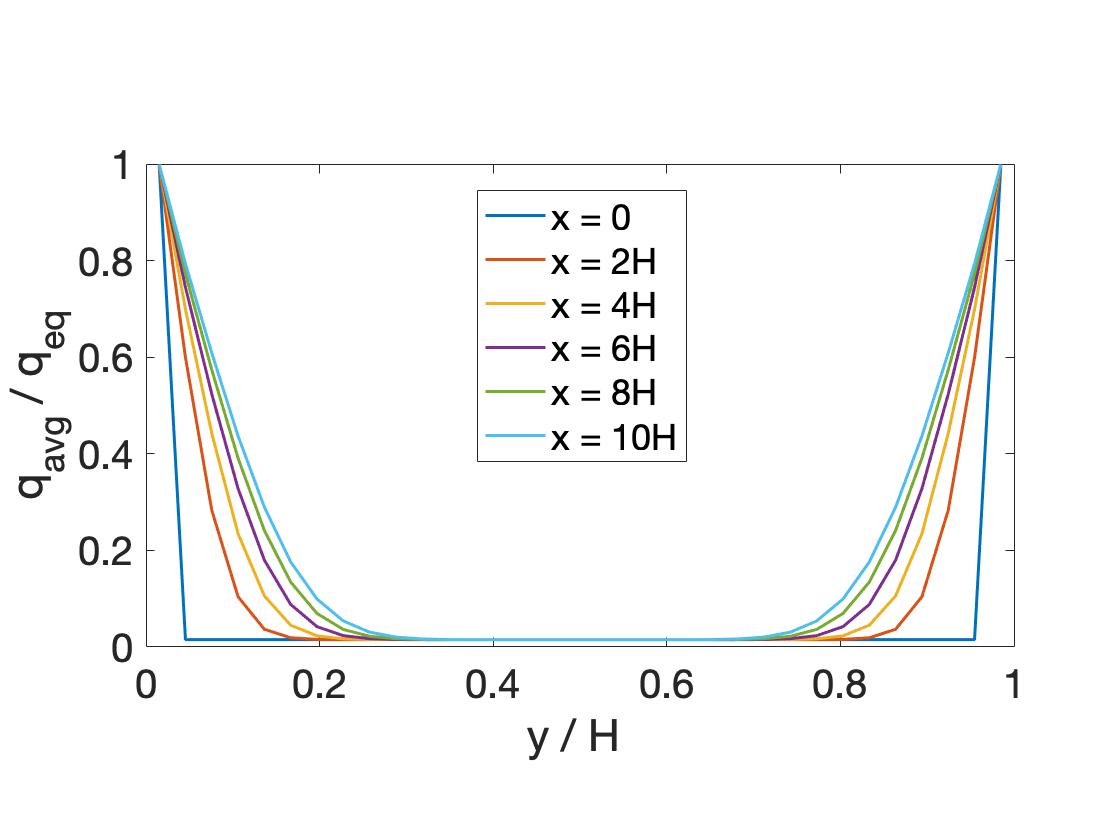}\label{fig:000}}\\
\subfigure[Case B, increase of the initial particle velocity distribution compared to A.]{
\includegraphics[trim=0cm 2cm 0cm 4cm,clip=true,width=0.44\textwidth]{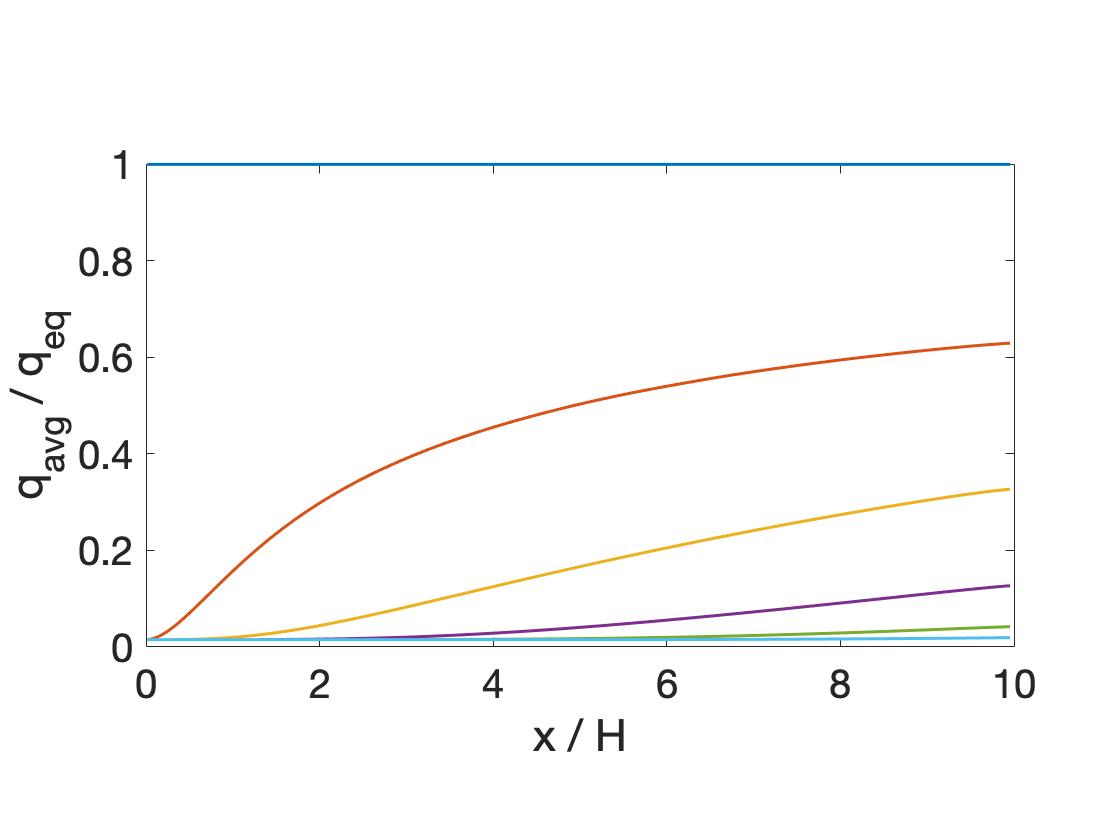}\qquad
\includegraphics[trim=0cm 2cm 0cm 4cm,clip=true,width=0.44\textwidth]{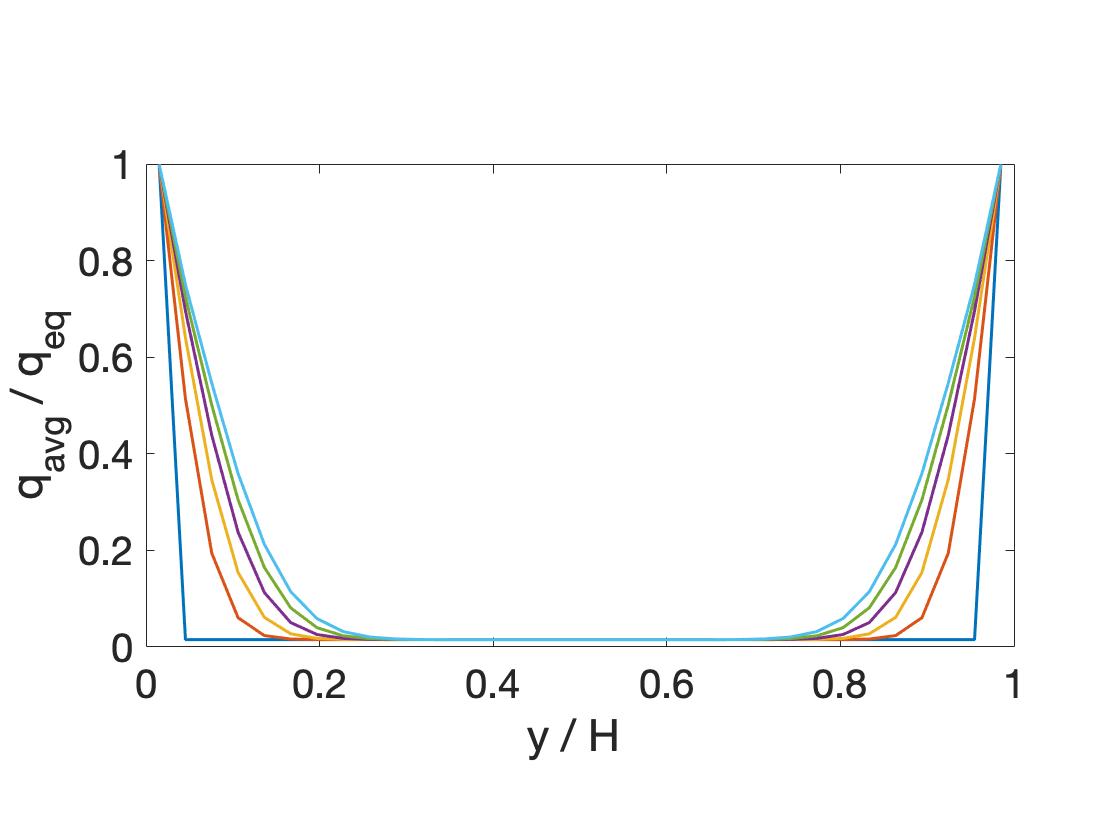}\label{fig:100}}\\
\subfigure[Case C, increase of the particle size distribution compared to A.]{
\includegraphics[trim=0cm 2cm 0cm 4cm,clip=true,width=0.44\textwidth]{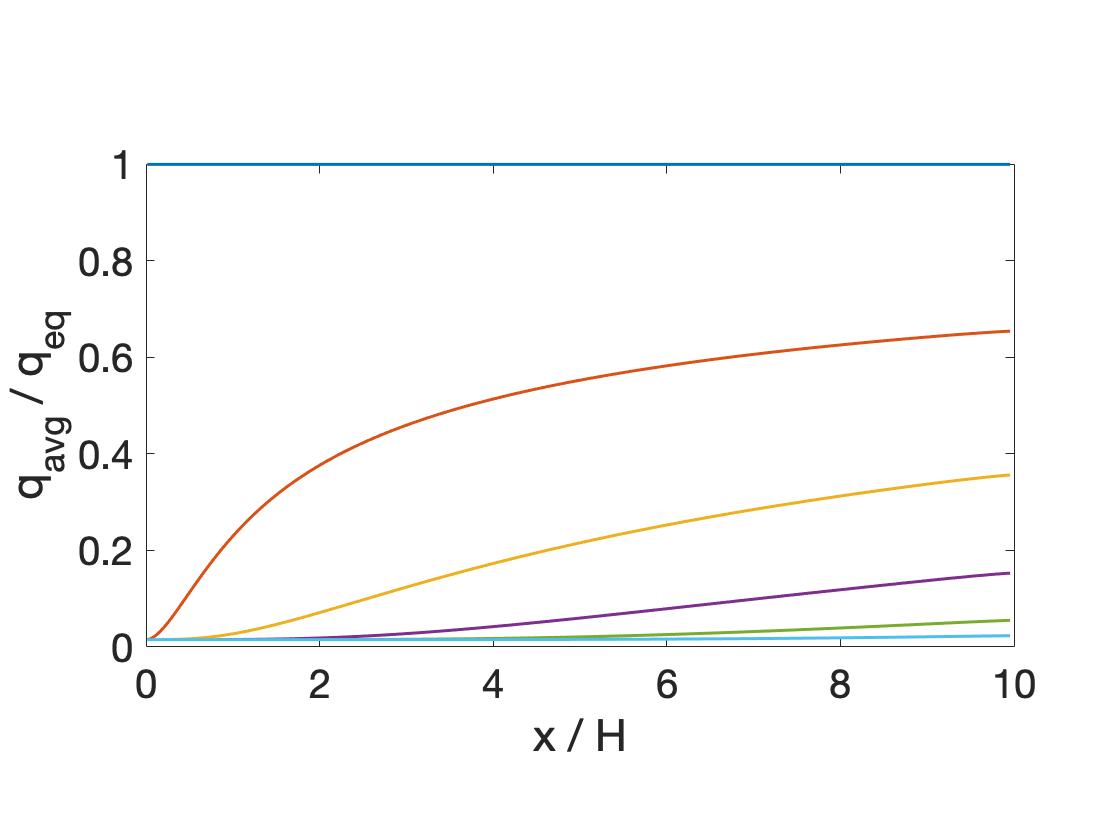}\qquad
\includegraphics[trim=0cm 2cm 0cm 4cm,clip=true,width=0.44\textwidth]{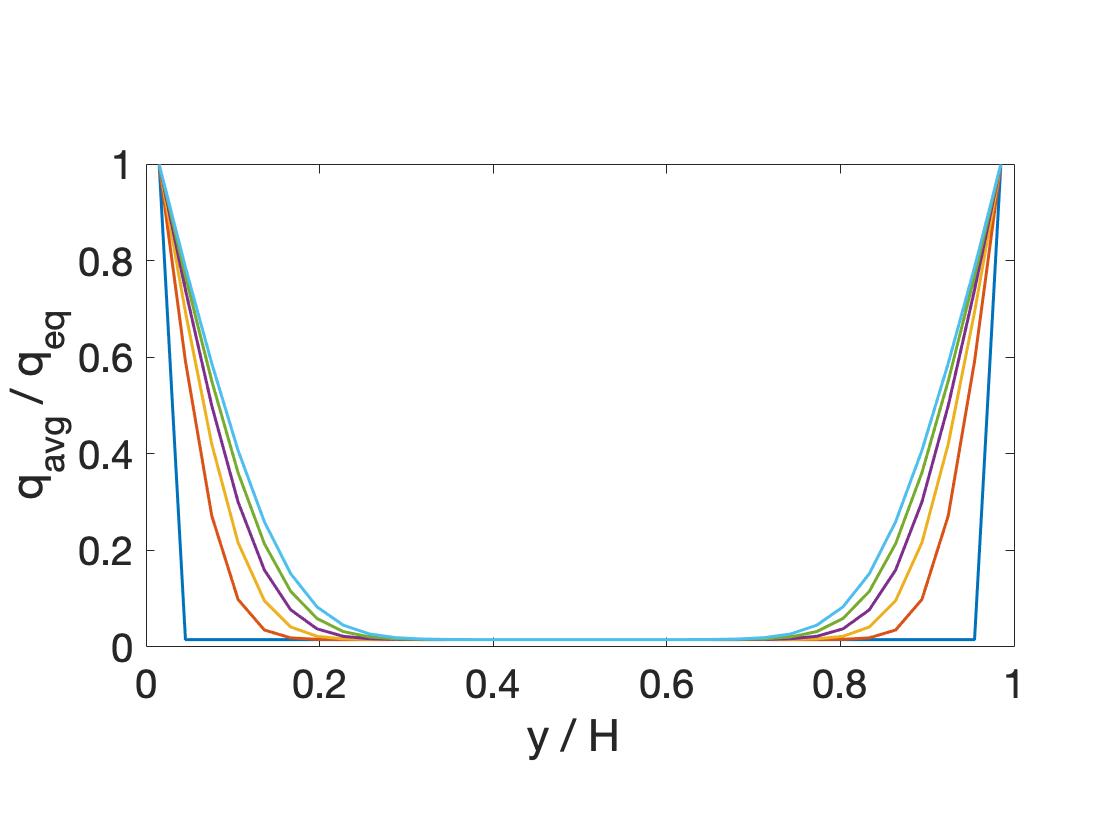}\label{fig:010}}\\
\subfigure[Case D, decrease of the charge diffusion coefficient compared to A.]{
\includegraphics[trim=0cm 2cm 0cm 4cm,clip=true,width=0.44\textwidth]{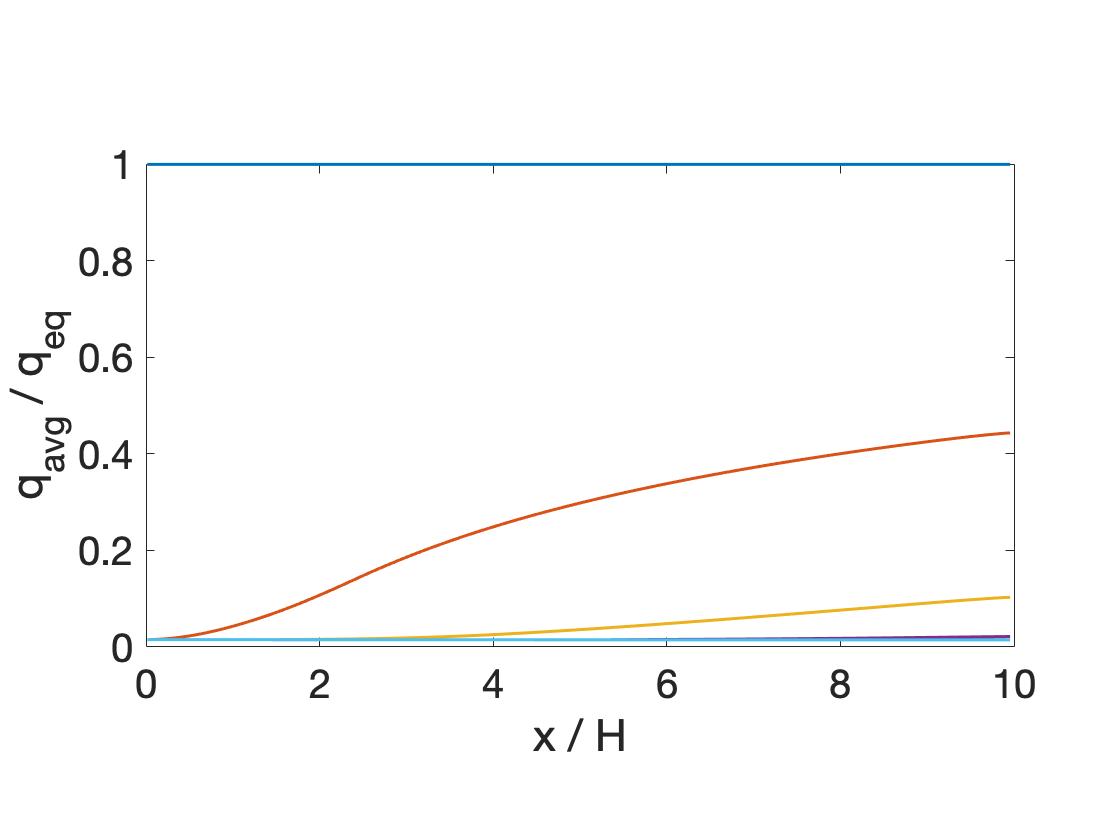}\qquad
\includegraphics[trim=0cm 2cm 0cm 4cm,clip=true,width=0.44\textwidth]{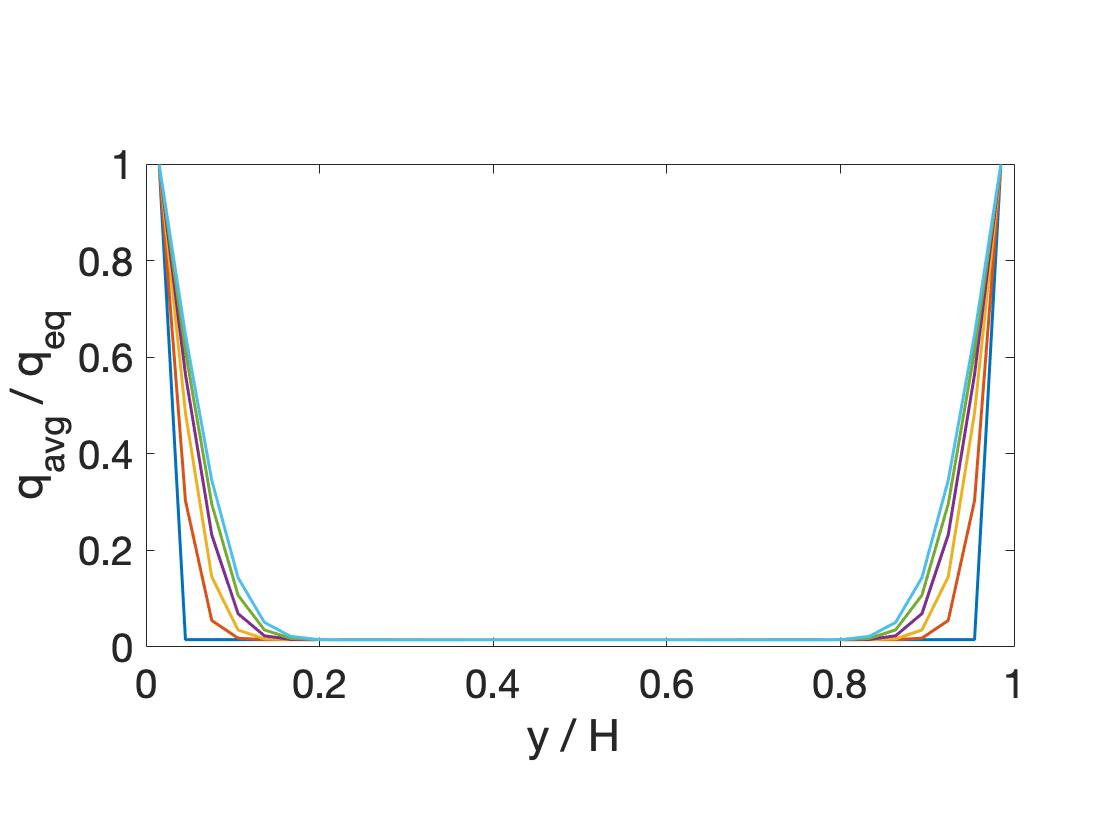}\label{fig:001}}
\caption{Streamwise (left column) and spanwise (right column) profiles of the average particle charge for case A (a) and the variation of one simulation parameter in cases B (b), C (c), and D (d).}
\label{fig:one}
\end{center}
\end{figure}

Next, we test the response of the solver to variations of individual simulation parameters.
Figure~\ref{fig:one} depicts the streamwise and spanwise profiles of the average particle charge for the cases~A to~D.
In each of the cases~B, C, and~D, one parameter is varied compared to case~A (cf.~Table~\ref{tab:table1}).

\begin{figure}[b]
\begin{center}
\includegraphics[trim=0cm 0cm 0cm 0cm,clip=true,width=0.44\textwidth]{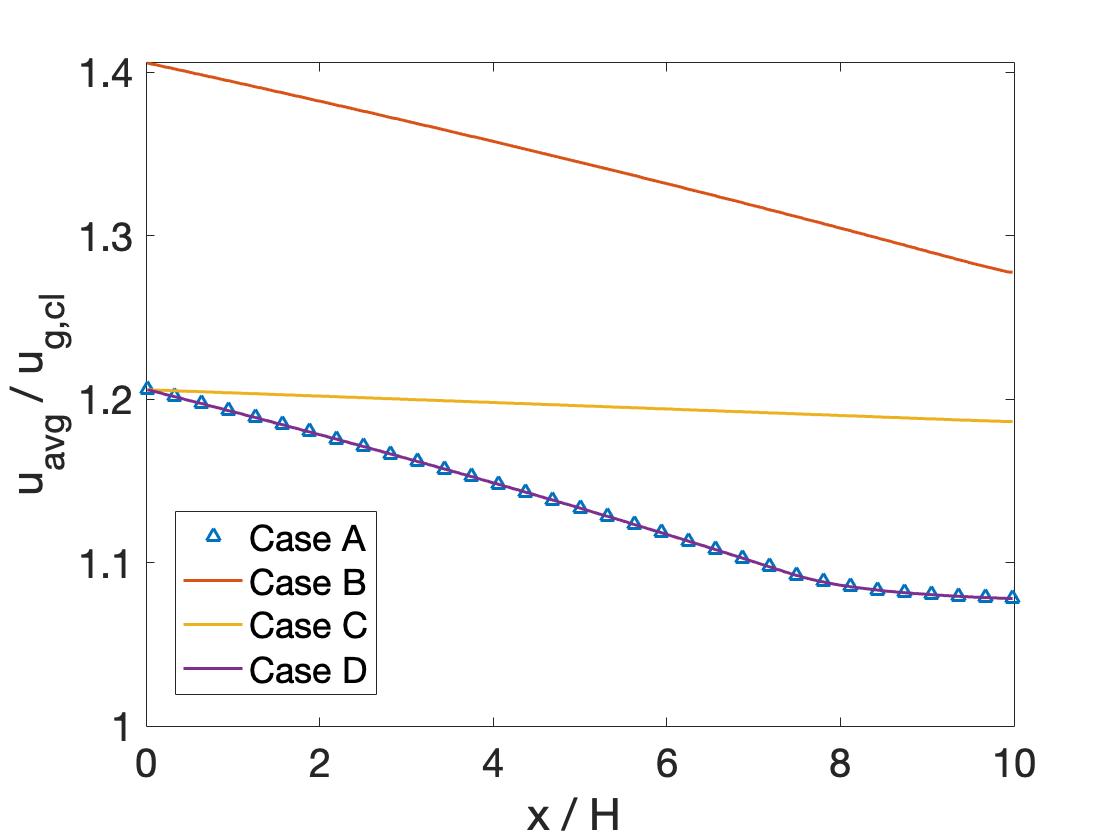}
\caption{Comparison of the average streamwise velocity of cases~A, B, C, and~D in the center.}
\label{fig:mean}
\end{center}
\end{figure}
Figure~\ref{fig:000} reiterates what was stated above for case A.
In the streamwise profiles (left column) one can observe how the flow is nearly uncharged at the inlet ($x=$\,0) and charges asymptotically towards the outlet ($x=10~H$).
The closer to the wall, the faster the particles charge while the lowest charge levels occur at the centreline ($y=0$).
The profile closest to the wall shows the charge boundary condition imposed on the flow.
Due to the negligible wall-nomal velocity, the cross-sectional profiles appear symmetrical.
Similarly, the right column illustrates the charge diffusion in the wall-normal direction.
Again, the profile located the closest to the inlet reflects the inlet and wall boundary conditions.
Then, in flow direction the charge migrates from the saturated walls towards the bulk of the flow.
In each cross-section in the downstream direction the average particle charge increases.
Apparently, the complete flow will asymptotically reach the equilibrium charge of the particles.
However, the domain in our study is not sufficiently long for the charge to reach the center of the channel.

Compared to case~A, in case~B (Figure~\ref{fig:100}) a powder flow of a higher initial velocity is computed.
As can be seen in both the stream- and spanwise profiles, the particles, in this case, charge slower than in case~A.
Also, a lower charge level is reached at the outlet.
Due to the higher streamwise velocity (Figure~\ref{fig:mean}), the particles spend less time traveling from the inlet to the outlet of the channel.
Thus, the charge has less time to migrate from the walls to the center.
In other words, due to the higher convective velocity, the resulting charge boundary layer is thinner.

Next, in Figure~\ref{fig:010} the results of case~C are plotted.
In this simulation, a larger particle size distribution is considered.
The profiles show that larger particles charge slower than the smaller particles of case~A.
This is related to the higher inertia, respecively Stokes number, of those particles through which they are less affected by the aerodynamic drag.
Thus, they travel faster through the domain than the particles of case~A, as it can be seen in Figure~\ref{fig:mean}.

The bottom row, Figure~\ref{fig:001}, shows case~D where the charge diffusion coefficient is reduced compared to case~A.
Also, in this case, the particles charge slower.
Similar to the above discussion regarding case~B, this effect is explained by the balance of charge motion perpendicular and parallel to the wall.
Due to the reduced diffusion the perpendicular charge transport is retarded which leads to a thinner charge boundary layer.

The above-discussed results confirm that the new numerical method responds as expected to variations of simulation parameters.
Further, the analysis in terms of Eulerian quantities suggests a non-dimensional quantity which expresses the balance of the convective and diffusive charge transport,  
\begin{equation}
\Phi=\frac{U L}{\nu_q} \, ,
\end{equation}
%
where $U$ is a characteristic particle flow velocity and $L$ is a characteristic length-scale, for example the channel width.
Analogous to the Reynolds, Nusselt, or Schmidt numbers, this new quantity indicates the thickness of the charge boundary layer.
For example, in the herein considered cases $\Phi$ is smaller for case~A than for cases~B and~C (where $U$ is higher) and case~D (where $\nu_q$ is higher) which exhibit thinner charge boundary layers.

\subsection{Cross-correlations of simulation parameters}

\begin{figure}[tb]
\begin{center}
\subfigure[Case E, increase of the initial particle velocity and particle size distribution compared to A.]
{\includegraphics[trim=0cm 2cm 0cm 4cm,clip=true,width=0.44\textwidth]{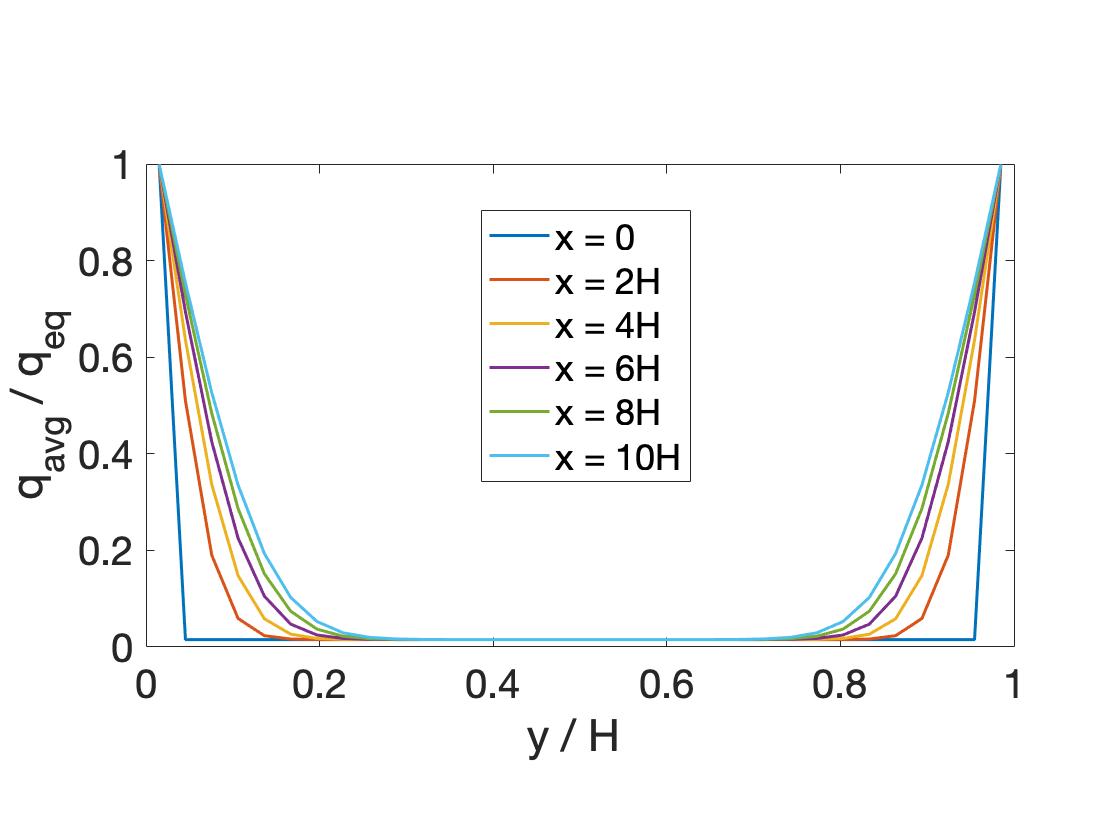}\label{fig:110}}\qquad
\subfigure[Case F, increase of the particle size distribution and decrease of the charge diffusion coefficient compared to A.]
{\includegraphics[trim=0cm 2cm 0cm 4cm,clip=true,width=0.44\textwidth]{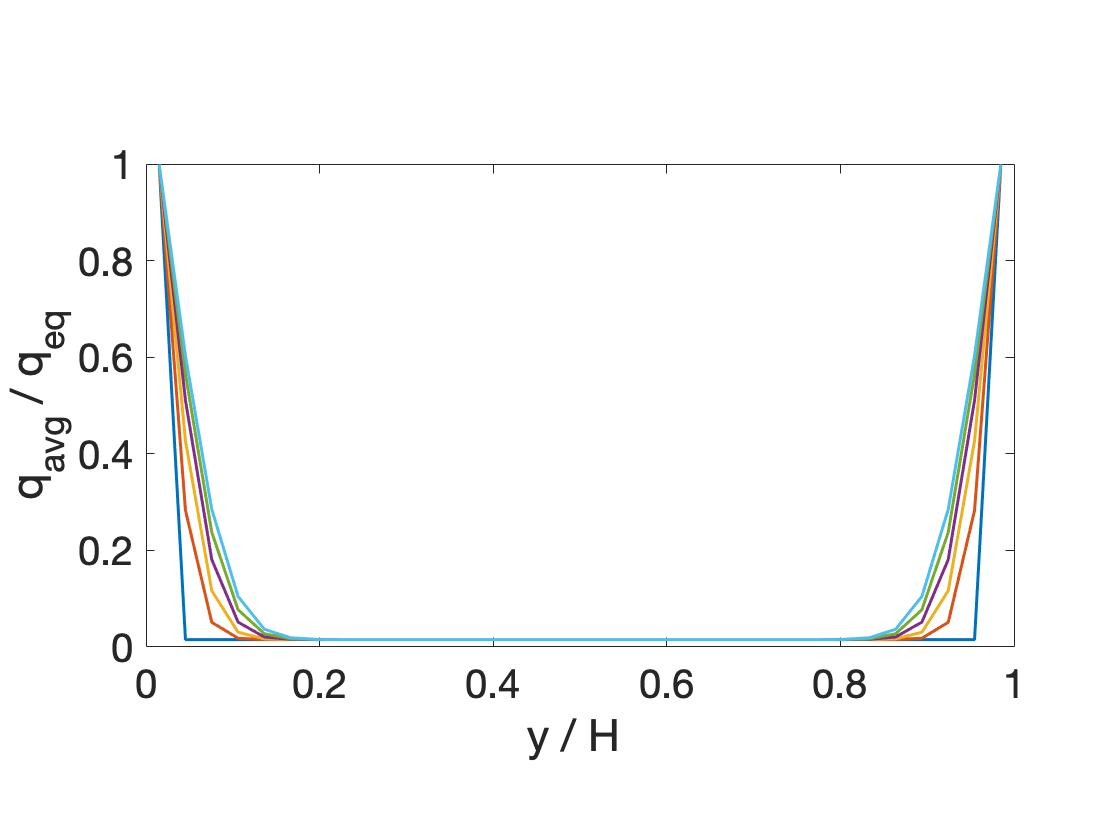}\label{fig:011}}\\
\subfigure[Case G, increase of the initial particle velocity distribution and decrease of the charge diffusion coefficient compared to A.]
{\includegraphics[trim=0cm 2cm 0cm 4cm,clip=true,width=0.44\textwidth]{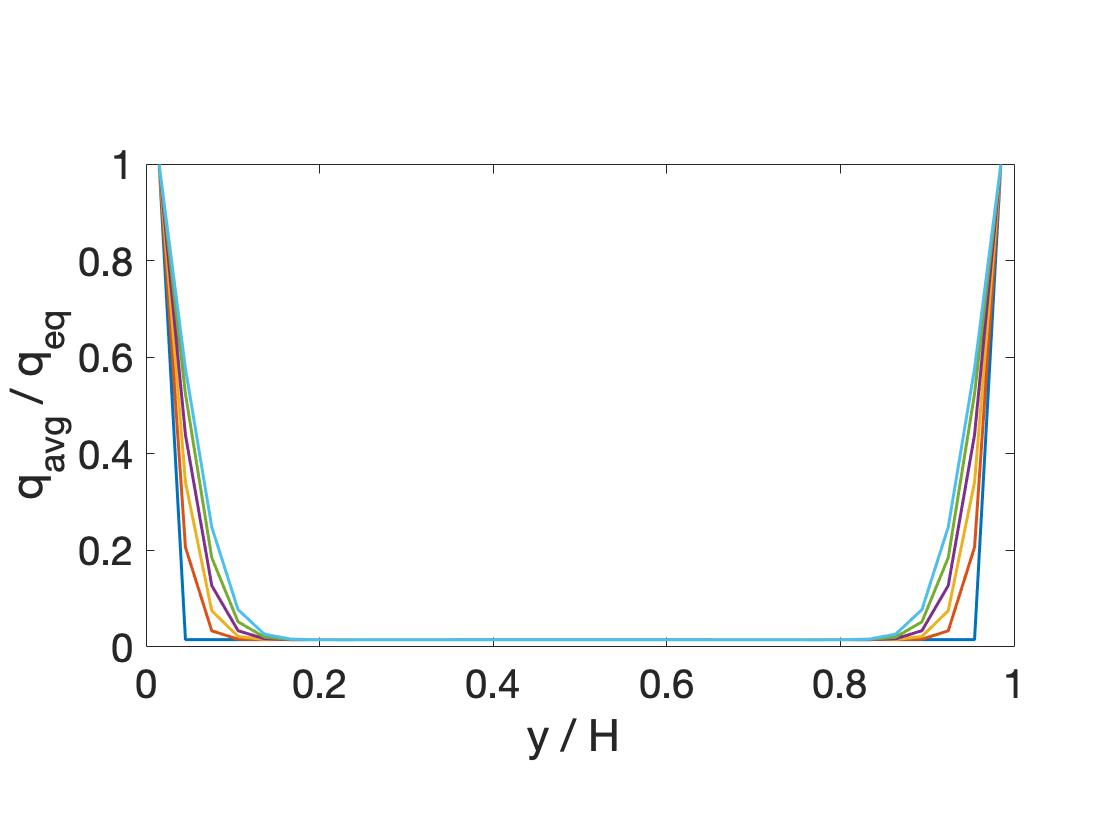}\label{fig:101}}\qquad
\subfigure[Case H, increase of the initial particle velocity and particle size distribution, and decrease of the charge diffusion coefficient compared to A.]
{\includegraphics[trim=0cm 2cm 0cm 4cm,clip=true,width=0.44\textwidth]{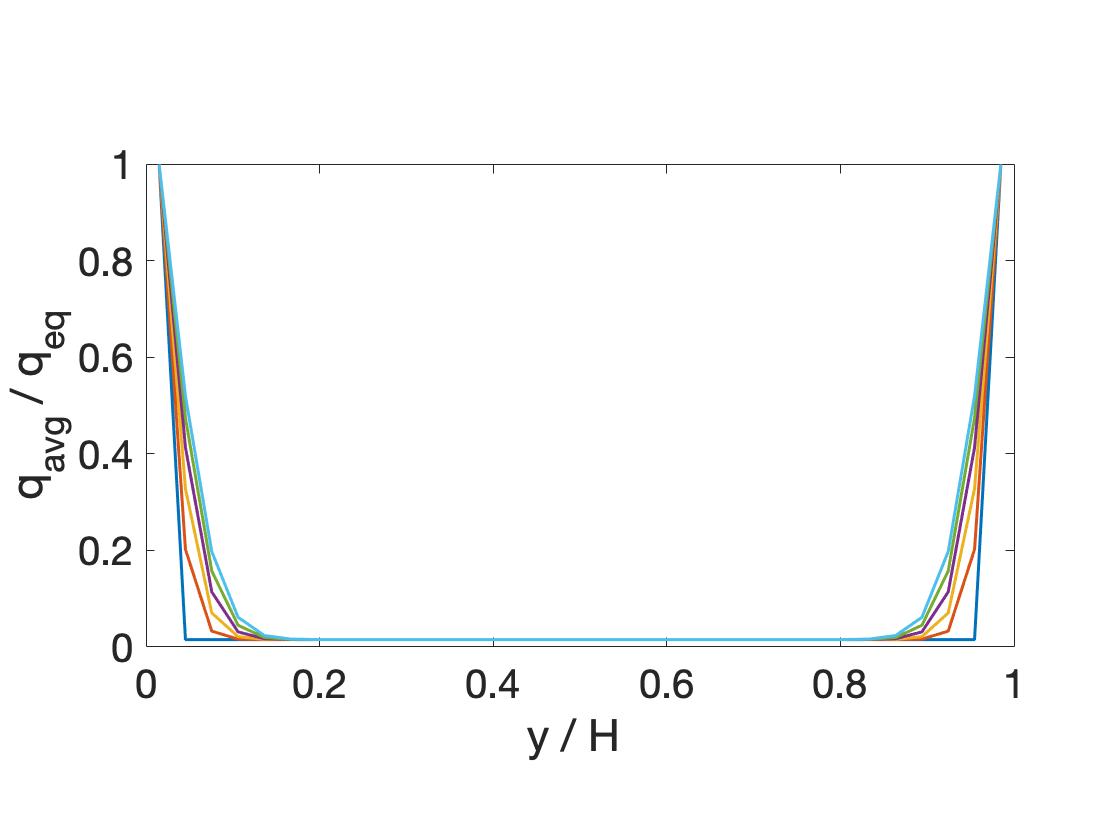}\label{fig:111}}
\caption{Spanwise profiles of the average particle charge when varying two or three parameters in cases E (a), F (b), G (c), and H (d).}
\label{fig:23}
\end{center}
\end{figure}

In this section cross-correlation effects of different simulation parameters are evaluated.
The resulting spanwise charge profiles when changing two or three parameters at the same time are presented in Figure~\ref{fig:23}.
In essence, the figures demonstrate that the effects of the individual parameters superimpose when several parameters are varied.

Thus, the particles charge the slowest in case~H where all parameters are varied, see Figure~\ref{fig:111}.
As revealed above, all parameter variations which are considered in our study lead individually to a slower charging.
Only slightly higher is the resulting particle charge of case~F in Figure~\ref{fig:011}.
Here the two-parameter variations which showed in Figure~\ref{fig:one} the strongest effect are applied, namely the increase of the particle size distribution and the decrease of the charge diffusion coefficient.

In both cases~E (Figure~\ref{fig:110} and~G (Figure~\ref{fig:011}) the initial particle velocity distribution is increased which showed in Figure~\ref{fig:100} to be of a rather low influence compared to the other parameters.
Further, in case~G the charge diffusion coefficient is decreased which leads to only slightly lower particle charges compared to case~E where the initial particle velocity is increased.
The results of these cases corroborate the significance of the balance of the convective and diffusive charge motion.
In other words, the charging of a given powder can be expected by the proposed non-dimensional number~$\Phi$.

\section{Conclusions}

This paper extends the DQMOM to compute the triboelectric charging of powder in an Eulerian framework.
The outstanding feature of this method is the ability to handle polydisperse particle size distributions.
The simulations of two-dimensional channel flows predicted the charge uptake of the particles at the walls, the convective charge transport in downstream and the diffusive charge transport in the wall-normal direction.
Moreover, the solver responded as expected to changes of the initial velocity and size distributions and the charge diffusion coefficient.
The ability to consider polydispersity in an Eulerian framework is a step towards the simulation of the charge build-up in technical flow facilities.
The next steps on the way to reach this goal are the extension of the algorithm from two to three dimensions, to model instationary flows, and to couple the solver to a CFD simulation of the gaseous phase.

\section*{Acknowledgments}

This project has received funding from the DECHEMA Max Buchner Research Foundation (grant No.~3736) and the European Research Council (ERC) under the European Union’s Horizon 2020 research and innovation programme (grant agreement No.~947606 PowFEct).

\section*{Data Availability Statement}
The data that support the findings of this study are available from the corresponding author upon reasonable request.

\appendix*
\section{Details of the mathematical derivation}
\label{math}

The quadrature approximation (equation~\ref{qadratureapprox}) is substituted to the left hand side equation~(\ref{eq:williams}) which yields
\begin{align}
\label{eq:lhs2}
\dfrac{\partial f}{\partial t} + \dfrac{\partial ({\bm u}f)}{\partial {\bm x}} &= \dfrac{\partial}{\partial t} \left( \sum^N_{n=1} w_n \delta(r-r_n) \delta({\bm u}-{\bm u}_n) \delta(q-q_n) \right) 
\nonumber \\
&+ \dfrac{\partial}{\partial {\bm x}} \left({\bm u}_n \sum^N_{n=1} w_n \delta(r-r_n) \delta({\bm u}-{\bm u}_n) \delta(q-q_n) \right) \, .
\end{align}
Using the product rule, this equation expands to 
\begin{align}
\label{eq:lhs3}
\dfrac{\partial f}{\partial t} + \dfrac{\partial ({\bm u}f)}{\partial {\bm x}} 
&= \sum^N_{n=1} \delta(r-r_n) \delta({\bm u}-{\bm u}_n) \delta(q-q_n) \left( \dfrac{\partial w_n}{\partial t} + \dfrac{\partial u_n w_n}{\partial {\bm x}} \right) 
\nonumber \\
&+ \sum^N_{n=1} w_n \delta'(r-r_n) \delta({\bm u}-{\bm u}_n) \delta(q-q_n) \left( -\dfrac{\partial r_n}{\partial t} - {\bm u}_n \dfrac{\partial r_n}{\partial {\bm x}} \right)
\nonumber \\
&+ \sum^N_{n=1} w_n \delta(r-r_n) \delta'({\bm u}-{\bm u}_n) \delta(q-q_n) \left( -\dfrac{\partial {\bm u}_n}{\partial t} - {\bm u}_n \dfrac{\partial {\bm u}_n}{\partial {\bm x}} \right)
\nonumber \\
&+ \sum^N_{n=1} w_n \delta(r-r_n) \delta({\bm u}-{\bm u}_n) \delta'(q-q_n) \left(  -\dfrac{\partial q_n}{\partial t} - {\bm u}_n \dfrac{\partial q_n}{\partial {\bm x}} \right) \, .
\end{align}
Now, the following form of source terms is introduced,
\begin{align}
\label{eq:lhs41a}
&\dfrac{\partial (w_n)}{\partial t} + \dfrac{\partial (w_n {\bm u}_n)}{\partial {\bm x}} = a_n
\\
\label{eq:lhs42a}
&\dfrac{\partial (w_n r_n)}{\partial t} + \dfrac{\partial (w_n {\bm u}_n r_n)}{\partial {\bm x}} = b_n
\\
\label{eq:lhs43a}
&\dfrac{\partial (w_n r_n {\bm u}_n)}{\partial t} + \dfrac{\partial (w_n {\bm u}_n r_n {\bm u}_n)}{\partial {\bm x}} = {\bm c}_n
\\
\label{eq:lhs44a}
&\dfrac{\partial (w_n r_n q_n)}{\partial t} + \dfrac{\partial (w_n {\bm u}_n r_n q_n)}{\partial {\bm x}} = d_n \, .
\end{align}
Combining and reordering equations~(\ref{eq:lhs42a}) and~(\ref{eq:lhs41a}) yields
\begin{equation}
\label{eq:lhs52}
w_n \left( \dfrac{\partial r_n}{\partial t} + {\bm u}_n \dfrac{\partial r_n}{\partial {\bm x}} \right) = b_n -r_n a_n \, .
\end{equation}
Similarly, combining equations~(\ref{eq:lhs43a}) and~(\ref{eq:lhs42a}) results in
\begin{equation}
\label{eq:lhs62}
w_n r_n \left( \dfrac{\partial {\bm u}_n}{\partial t} + {\bm u}_n \dfrac{\partial {\bm u}_n}{\partial {\bm x}} \right) = {\bm c}_n - {\bm u}_n b_n \, ,
\end{equation}
and equations~(\ref{eq:lhs44a}) and~(\ref{eq:lhs42a}) in
\begin{equation}
\label{eq:lhs72}
w_n r_n \left( \dfrac{\partial {\bm u}_n}{\partial t} + {\bm u}_n \dfrac{\partial {\bm u}_n}{\partial {\bm x}} \right) = d_n - q_n b_n \, .
\end{equation}

The substitution of equations~(\ref{eq:lhs41a}), (\ref{eq:lhs52}), (\ref{eq:lhs62}), and (\ref{eq:lhs72}) in equation~(\ref{eq:lhs3}) and reordering the terms leads to
\begin{align}
\label{eq:lhs82}
\dfrac{\partial f}{\partial t} + \dfrac{\partial ({\bm u}f)}{\partial {\bm x}} 
&= \sum^N_{n=1} \left( \delta(r-r_n) \delta({\bm u}-{\bm u}_n) \delta(q-q_n) + r_n \delta'(r-r_n) \delta({\bm u}-{\bm u}_n) \delta(q-q_n) \right) a_n
\nonumber \\
&- \sum^N_{n=1} \left( \delta'(r-r_n) \delta({\bm u}-{\bm u}_n) \delta(q-q_n) - \dfrac{{\bm u}_n}{r_n} \delta(r-r_n) \delta'({\bm u}-{\bm u}_n) \delta(q-q_n) \right) b_n
\nonumber \\
&- \sum^N_{n=1} \left( \dfrac{1}{r_n} \delta(r-r_n) \delta'({\bm u}-{\bm u}_n) \delta(q-q_n) - \dfrac{q_n}{{\bm u}_n r_n} \delta(r-r_n) \delta({\bm u}-{\bm u}_n) \delta'(q-q_n)  \right) {\bm c}_n
\nonumber \\
&- \sum^N_{n=1} \left( \dfrac{1}{{\bm u}_n r_n}  \delta(r-r_n) \delta({\bm u}-{\bm u}_n) \delta'(q-q_n) \right) d_n \, .
\end{align}

Thereafter, the moment transformation is applied to equation~(\ref{eq:lhs82}).
The first term on the right hand side of equation~(\ref{eq:lhs82}) gives
\begin{align}
\label{eq:lhs91}
\int r^k u_1^l u_2^m u_3^p q^s &\Big( \delta(r-r_n) \delta({\bm u}-{\bm u}_n) \delta(q-q_n) \nonumber \\
&+ r_n \delta'(r-r_n) \delta({\bm u}-{\bm u}_n) \delta(q-q_n) \Big) a_n dr du_1 du_2 du_3 dq \nonumber \\
&= (1-k) r^k_n u^l_{1,n} u^m_{2,n} u^p_{3,n} q^s_n a_n \, ,
\end{align}
the second term
\begin{align}
\label{eq:lhs92}
\int r^k u_1^l u_2^m u_3^p q^s &\Big( \delta'(r-r_n) \delta({\bm u}-{\bm u}_n) \delta(q-q_n) \nonumber \\
&- \dfrac{{\bm u}_n}{r_n} \delta(r-r_n) \delta'({\bm u}-{\bm u}_n) \delta(q-q_n) \Big) b_n dr du_1 du_2 du_3 dq\nonumber \\
&= (-k+l+m+p) r^{k-1}_n u^l_{1,n} u^m_{2,n} u^p_{3,n} q^s_n b_n \, ,
\end{align}
the third term
\begin{align}
\label{eq:lhs93}
\int r^k u_1^l u_2^m u_3^p q^s &\Big( \dfrac{1}{r_n} \delta(r-r_n) \delta'({\bm u}-{\bm u}_n) \delta(q-q_n) \nonumber \\
&- \dfrac{q_n}{{\bm u}_n r_n} \delta(r-r_n) \delta({\bm u}-{\bm u}_n) \delta'(q-q_n)  \Big) {\bm c}_n dr du_1 du_2 du_3 dq \nonumber \\
&= \left( l u^{-1}_{1,n} c_{1,n} + m u^{-1}_{2,n} c_{2,n} + p u^{-1}_{3,n} c_{3,n} \right) r^{k-1}_n u^l_{1,n} \, ,
\end{align}
and the fourth term
\begin{align}
\label{eq:lhs94}
\int r^k u_1^l u_2^m u_3^p q^s & \left( \dfrac{1}{{\bm u}_n r_n}  \delta(r-r_n) \delta({\bm u}-{\bm u}_n) \delta'(q-q_n) \right) d_n dr du_1 du_2 du_3 dq \nonumber \\
&= -q r^{k-1}_n u^{l-1}_{1,n} u^{m-1}_{2,n} u^{p-1}_{3,n} q^{s-1}_n {\bm c}_n \, .
\end{align}

Finally, substituting equations~(\ref{eq:lhs91}) to (\ref{eq:lhs94}) in equation~(\ref{eq:lhs82}) yields
\begin{align}
\int r^k u_1^l u_2^m u_3^p q^s &\left( \dfrac{\partial f}{\partial t} + \dfrac{\partial {\bm u} f}{\partial {\bm x}} \right) dr du_1 du_2 du_3 dq =
\nonumber \\
& \sum^N_{n=1} (1-k) r^k_n u^l_{1,n} u^m_{2,n} u^p_{3,n} q^s_n a_n
\nonumber \\
+ & \sum^N_{n=1} (-k+l+m+p) r^{k-1}_n u^l_{1,n} u^m_{2,n} u^p_{3,n} q^s_n b_n
\nonumber \\
+ & \sum^N_{n=1} \left( l u^{-1}_{1,n} c_{1,n} + m u^{-1}_{2,n} c_{2,n} + p u^{-1}_{3,n} c_{3,n} \right) r^{k-1}_n u^l_{1,n} u^m_{2,n} u^p_{3,n} q^s_n
\nonumber \\
+ & \sum^N_{n=1} -s r^{k-1}_n u^{l-1}_{1,n} u^{m-1}_{2,n} u^{p-1}_{3,n} q^s_n {\bm c}_n \nonumber \\
+ & \sum^N_{n=1} -s r^{k-1}_n u^{l-1}_{1,n} u^{m-1}_{2,n} u^{p-1}_{3,n} q^{s-1}_n d_n \, .
\end{align}
To sum up, equation~\ref{eq:lhs9} represents the left-hand side of equation~(\ref{eq:lhs9}) after the quadrature-based approximation and the momentum transformation.

\bibliography{publications}

\end{document}